\documentclass[a4paper,fleqn]{cas-sc}
\usepackage{amsmath,amssymb,amsfonts,amsthm}
\usepackage{graphicx}
\usepackage{dcolumn}
\usepackage{bm}
\usepackage[dvipsnames]{xcolor}
\usepackage[normalem]{ulem}
\usepackage[bf]{subfigure}
\usepackage{float}
\usepackage[T1]{fontenc}
\usepackage{hyperref}
\usepackage{adjustbox}
\usepackage[numbers,sort&compress]{natbib}
\usepackage{pdfpages}

\newcommand{\novo}[1]{{\color{black}#1}}

\begin{document}
\let\WriteBookmarks\relax
\def\floatpagepagefraction{1}
\def\textpagefraction{.001}

\shorttitle{Sleep stages in the complexity--weighted-entropy plane}
\shortauthors{de Lucas et al.}

\title[mode=title]{Characterizing sleep stages through the complexity--weighted-entropy plane in human intracranial data and in a whole-brain model}

\author[ufal,ifisc]{Helena Bordini de Lucas}\cormark[1]
\ead{helena.bordini@fis.ufal.br}

\author[idibaps]{Leonardo Dalla~Porta}[orcid=0000-0003-3104-288X]
\author[neuropsi]{Alain Destexhe}[orcid=0000-0001-7405-0455]
\author[idibaps,icrea]{Maria V. Sanchez-Vives}[orcid=0000-0002-8437-9083]
\author[ufal]{Osvaldo A. Rosso}[orcid=0000-0002-1288-2528]
\author[ifisc]{Cláudio R. Mirasso}[orcid=0000-0003-2980-7038]
\author[ufal]{Fernanda Selingardi Matias}[orcid=0000-0002-5629-6416]

\cortext[1]{Corresponding author}

\affiliation[ufal]{organization={Instituto de F\'{\i}sica, Universidade Federal de Alagoas},
            country={Brazil}}

\affiliation[ifisc]{organization={Instituto de F\'{\i}sica Interdisciplinar y Sistemas Complejos, Universitat de les Illes Balears},
            country={Spain}}

\affiliation[idibaps]{organization={Institute of Biomedical Research August Pi Sunyer (IDIBAPS), Systems Neuroscience},
            city={Barcelona},
            country={Spain}}

\affiliation[neuropsi]{organization={Paris-Saclay University, CNRS, Paris-Saclay Institute of Neuroscience (NeuroPSI)},
            city={Saclay},
            country={France}}

\affiliation[icrea]{organization={ICREA},
            city={Barcelona},
            country={Spain}}

\begin{abstract}
Approaches based on symbolic information, such as entropy and complexity, have shown that sleep reorganizes the temporal structure of brain activity. However, most of this evidence comes from scalp EEG and group averages. In this work, we characterized sleep stages in intracranial recordings from 65 patients during wakefulness and the REM, N2, and N3 stages, using weighted permutation entropy (WPE) and the statistical complexity measure (SCM). In the complexity--entropy plane, the states formed a clear trajectory from wakefulness to REM, N2, and N3, with decreasing entropy and increasing complexity as sleep deepened. This organization was observed individually in most patients and was recovered more consistently by WPE than by conventional permutation entropy. The trajectory, however, was not the same throughout the cortex. All regions shifted in the same direction, but the occipital cortex followed a much shorter path than the parietal cortex. This result reflects the spatial heterogeneity of the brain during sleep and shows that cortical regions differ not only in the position they occupy in the plane, but also in the path they follow between states. Finally, a whole-brain model based on the human connectome showed that increasing neuronal adaptation shifts the simulated dynamics in the same direction observed in the empirical data, pointing to adaptation as one of the candidate mechanisms involved in this reorganization.
\end{abstract}


\maketitle

\section{Introduction}
\label{Sec-Introduction}

The brain can be studied as a dynamical system composed of many variables that continuously reconfigure over time, whose evolution can be described through time series. In the study of sleep, the electroencephalogram (EEG)~\cite{rechtschaffen1968manual} constitutes the main basis for this description. It is well established that the sleep--wake cycle comprises three main states, wakefulness (Wake), non-rapid eye movement sleep (NREM), and rapid eye movement sleep (REM), which alternate throughout the night~\cite{Dement1957}. NREM is further subdivided into stages N1, N2, and N3, the latter corresponding to slow-wave sleep~\cite{AASM}. The tools used to characterize these states depend on the signal property one wishes to investigate, some describe its spectral content~\cite{aeschbach1993allnight, prerau2017sleep}, others the amplitude of its fluctuations~\cite{agnew1967amplitude, feinberg1978period}, and others the temporal organization of the time-series values themselves, as occurs in methods based on ordinal patterns~\cite{bandt2017new, nicolaou2011permutation}. \novo{These approaches belong to a broader family of nonlinear, multiscale, and network-based methods used to characterize brain dynamics~\cite{Kralj2026, Gao2021}.}

Permutation entropy (PE), proposed by Bandt and Pompe~\cite{BandtPompe2002}, transforms segments of the series into symbols defined by the relative ordering of their elements and constructs, from their occurrences, a probability distribution linked to the dynamics of the series. Calculated over this distribution, PE quantifies the diversity of ordinal patterns and has been widely applied to physiological signals~\cite{Lotfi2020b, Rosso06, CARLOS2025134955}, including EEG during sleep~\cite{bandt2017new, nicolaou2011permutation, mateos2021using, hou2021changes, guo2021multi}. Because it considers only the ordering of the values, however, PE treats patterns with very different amplitudes in the same way. This limitation is particularly relevant in sleep, whose NREM stages are marked by large-amplitude events, such as slow waves and K-complexes~\cite{massimini2004sleep, cash2009human}. Weighted permutation entropy (WPE)~\cite{WPE_2013} circumvents this problem by incorporating into the pattern distribution not only the relative ordering of the values but also the amplitude of their fluctuations, and weighted variants have already been shown to outperform conventional PE in sleep-stage classification using scalp EEG~\cite{duarte2025statistical}.

\novo{Entropy measures unpredictability, but the same decrease in entropy may result from either a
simplification of the dynamics or the emergence of temporal structure. White noise illustrates this
distinction: it is highly unpredictable but dynamically trivial. The statistical complexity measure (SCM)
can therefore complement entropy by distinguishing between these cases}~\cite{Lopez1995, RossoPRL, lamberti2004intensive}. The SCM weights the entropy of the distribution by its distance from equiprobability, also capturing the degree of structure present in the ordinal patterns. The joint analysis of the two quantities in the so-called complexity--entropy plane~\cite{RossoPRL} thus makes it possible to ask not only whether a state exhibits higher or lower entropy, but how the states are organized in this space and how they shift during transitions between them.

This representation has already been applied to sleep. Studies using ordinal quantifiers in scalp EEG and animal recordings consistently report that the transition to NREM sleep is accompanied by a reduction in entropy, particularly pronounced in deep sleep, whereas Wake and REM exhibit high and similar values~\cite{mateos2021using, gonzalez2019decreased, mateos2018measures}. \novo{Entropy-based measures have also revealed state-dependent and spatially localized EEG alterations in clinical populations, such as children with epilepsy and comorbid ADHD~\cite{Cai2026} and patients with Parkinson's disease~\cite{Davoodi2025}.} Three questions, however, remain open. First, scalp EEG reflects the spatial mixing of widespread cortical sources, so it is unclear whether the position of each state in the plane is a property of local cortical dynamics or a byproduct of this mixing. Second, results are usually reported as population averages, without verifying whether the same organization is maintained when each participant is treated as an independent unit of observation. Third, the geometry observed in the plane has rarely been related to a neurodynamic process capable of generating it.

Intracranial EEG (iEEG) recordings provide a direct way to address the first two questions, as they access cortical activity in a localized manner and allow the analysis to be resolved by patient and cortical region. This latter point is particularly relevant because sleep is not spatially homogeneous, that is, different cortical regions may exhibit distinct degrees of synchronization and slow-wave recruitment during the same stage~\cite{Nir2011NREM, bernardi2018local, siclari2017local}, and it is precisely this heterogeneity that scalp spatial mixing and group averages tend to erase. Recent studies have applied information quantifiers to intracranial recordings from the same dataset used here~\cite{guisande2024neural, guisande2024renyi, jeantin2026atlas, kalamangalam2025gradients}, describing variations in entropy and complexity across stages and cortical regions, and entropy--complexity planes have even been employed to compare groups of participants within each state~\cite{guisande2024renyi}. These studies, however, compared values across recordings or regions, without treating sleep states as a trajectory in the complexity--entropy plane, without testing whether this ordering is repeated within each patient, and without asking whether different regions follow distinct paths between states.

In this work, we characterize sleep stages in the plane formed by WPE and SCM, applied to iEEG recordings from 65 patients obtained from an open-access database, the \href{https://mni-open-ieegatlas.research.mcgill.ca}{MNI Open iEEG Atlas}~\cite{Frauscher2018, frauscher2018high, von2020human}, in the Wake, REM, N2, and N3 states, with analyses at the global, individual, and regional levels. We complement this characterization with a whole-brain model composed of MF-AdEx populations coupled through a human structural connectome, in which a single biophysical parameter, the level of activity-dependent neuronal adaptation, can be varied in a controlled manner~\cite{DallaPorta2025WBM}. Because adaptation is regulated by neuromodulatory systems that change throughout the sleep--wake cycle~\cite{mccormick2020neuromodulation}, this comparison makes it possible to assess whether a plausible cellular mechanism is compatible with the geometry observed in the plane.

We find that the four states follow the trajectory Wake~$\rightarrow$~REM~$\rightarrow$~N2~$\rightarrow$~N3, with a progressive reduction in WPE and an increase in SCM. This ordering is not an artifact of population averaging, it is maintained when each patient is treated as a unit of observation and is recovered more consistently by WPE than by conventional PE, extending to intracranial recordings the advantage of amplitude weighting observed in scalp EEG. The regional analysis further shows that the plane is not traversed in the same way throughout the cortex, with the occipital and parietal trajectories differing the most, reflecting the spatial heterogeneity of sleep. Finally, increasing neuronal adaptation in the model reproduces the same direction of displacement, connecting the statistical characterization of cortical dynamics during sleep to a possible mechanistic interpretation of transitions between states.

\section{Information Theory Quantifiers: Weighted Permutation Entropy and Statistical Complexity Measure}
\label{Sec:Information-quantifiers}

Consider $P$ a probability distribution of $N$ possible events associated with probabilities $p_1, p_2, \dots, p_N$, with $0 \leq p_i \leq 1$ and $\sum_{i = 1}^{N}p_i = 1$. Shannon Entropy, $S(P)$, which quantifies the average of the information contained in $P$, is given by~\cite{shannon1948}:

\begin{equation}
S(P) = - \sum_{i=1}^N p_i \log_2{(p_i)}.
\label{eq:ShannonEntropy1}
\end{equation}

From Eq.~\eqref{eq:ShannonEntropy1}, the system is completely predictable when $S(P) = 0$. On the other hand, when all $N$ possible states are equally likely, the entropy reaches its maximum value, $S(P)_{\max} = \log_2(N)$, making the system maximally unpredictable.

\subsection{Weighted Permutation Entropy (WPE)}

Permutation Entropy (PE) is a natural technique used to analyze the ordinal patterns of neighboring elements within a time series, constructing a probability density function (PDF) that reflects the underlying time series dynamics. The original formulation of PE, introduced by Bandt and Pompe~\cite{BandtPompe2002}, considers only the relative order of values in the time series, without accounting for their actual amplitudes. A natural extension of this method is Weighted Permutation Entropy (WPE), which integrates amplitude information into the construction of the PDF~\cite{WPE_2013}.

\begin{figure}
  \centering 
  \includegraphics[trim=0 52 0 0, clip, width=0.45\linewidth]{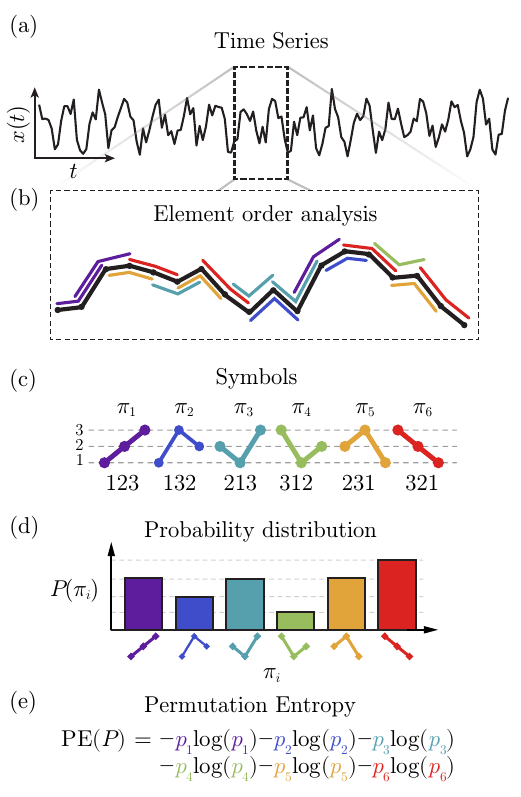}  
  \caption{Illustration of the Bandt-Pompe methodology for transforming a time series into a probability distribution using a symbolic approach. (a) Representative time series. (b) Segmentation process for $D = 3$. (c) The six possible permutation patterns, where $N = D!$. (d) Probability distribution corresponding to the permutation patterns and their associated weights.}
  \label{fig:symbol}
\end{figure}

Consider a time series $\mathcal{X}(t)$ of length $M$, represented in Fig.~\ref{fig:symbol}(a). The first step of the approach consists of dividing this series into overlapping vectors with $D$ (embedding dimension) elements, defined by:

\begin{equation}
{\vec{Y}}_{j} = (x_j, x_{j+\tau}, x_{j+2\tau}, ~\dots~, x_{j+(D-1)\tau} )\ ,
\label{eq:D_vectors}
\end{equation}

where $\tau$ is the embedding time or time delay ($\tau \in \mathbb{N}, \tau \geq 1$) and $j$ varies from 1 to $M - (D-1)\tau$. This segmentation process can be seen in Fig.~\ref{fig:symbol}(b), where a subset of the time series is highlighted to exemplify the construction of the vectors ${\vec{Y}}_j$.

Each vector is then converted into a permutation symbol ${\pi}_i$ based on the relative ordering of its elements, where $j$ indexes the symbolic pattern extracted from the original time series. The ordering is defined by assigning to each component of the vector $\vec{Y}_j$ its rank among the $D$ components:

\begin{equation}
\pi_i = [r_1, r_2, \dots, r_D] \ ,
\label{eq:order-componentes}
\end{equation}

where each $r_k$ represents the position (or rank) of the $k$-th element of $\vec{Y}_j$ in the ascending ordering of its values, with $1$ assigned to the smallest and $D$ to the largest value of magnitude, e.g. $\vec{Y}_1 = ~(3,5,1)   \mapsto \pi_5 = [231]$,  since the value $3$ is the second smallest, $5$ is the largest (third), and $1$ is the smallest (first) among the elements of the vector.

This correspondence between the vectors ${\vec{Y}}_j$ and their respective symbols ${\pi}_i$ is shown in Fig.~\ref{fig:symbol}(c), where all possible permutation patterns for $D=3$ are presented. Since there are $N = D!$ possible distinct orderings of the elements in each vector, there are exactly six distinct symbols ($N = 3! = 6$) representing all possible configurations.

Each vector ${\vec{Y}}_j$ has an associated weight $w_j$, based on its variance,

\begin{equation}
w_j = \frac{1}{D}\sum_{k = 1}^{D}\left(x_{j+(k-1)\tau} - \overline{X}_j\right)^2, 
\label{eq:variance}
\end{equation}

where $\overline{X}_j$ stands for the mean value of the embedded vector $\vec{Y}_j$. 

Then the relative frequencies $p_i$ of each symbol $\pi_i$ is calculated as:

\begin{equation}
p_i =\frac{\sum_{j\leq M-(D-1)\tau}\delta_{ij}w_j}{\sum_{j\leq M-(D-1)\tau}w_j},
\label{eq:PDF_WPE}
\end{equation}

where $\delta_{ij}$ selects if $\vec{Y}_j$ is of symbol type $\pi_i$ 
\begin{equation}
\delta_{ij} =
   \begin{cases}
    1 & \text{if } \vec{Y}_j \text{ is of symbol type } \pi_i \\
    0 & \text{otherwise.}
    \end{cases}
\end{equation}

In Fig.~\ref{fig:symbol}(d), this process is represented as a histogram, where each bar corresponds to the relative frequency $p_i$ of each symbol $\pi_i$ identified in Fig.~\ref{fig:symbol}(c), along with their respective weights.

Finally, normalized Weighted Permutation Entropy is calculated based on this probability distribution $P$ according to the equation:

\begin{equation}
\mathrm{WPE} = -  \frac{1}{\log_2 D!}\sum_{i=1}^{D!} p_i \log_2(p_i).
\label{eq:W_Permutation_Entropy}
\end{equation}

This metric quantifies the uncertainty in the distribution of permutation patterns. A uniform distribution, where all probabilities are equally likely, indicates a highly unpredictable dynamics in the system ($\mathrm{WPE} \to 1$), whereas a distribution concentrated in a few symbols suggests more regular and predictable behavior ($\mathrm{WPE} \to 0$).

\subsection{Statistical Complexity Measure (SCM)}  

Between the two extremes of entropy (0 and 1) lie dynamical systems of great interest, as many natural processes are governed by dynamics that fall within this intermediate range. Since the ordinal structures present in a process are not fully captured by entropy measures alone, it becomes necessary to use statistical or structural complexity measures to obtain a more detailed characterization of the underlying dynamics represented by the time series~\cite{Lopez1995}.

Statistical Complexity allows for the identification of essential details of the dynamics and, more importantly, characterizes the correlational structure of the orderings present in time series~\cite{Lopez1995}. A fundamental property of this measure is that its value must be zero at the two extremes of entropy and reach a maximum for intermediate entropy values. From a qualitative perspective, this behavior can be interpreted as a measure of non-triviality rather than mere unpredictability. For instance, a time series may be highly unpredictable, as in the case of white noise, while still representing a trivial system in this framework, since the Statistical Complexity is zero in such cases.\color{black}

The Statistical Complexity Measure used here is defined by~\cite{RossoPRL, lamberti2004intensive, martin2006generalized}:

\begin{equation}
\label{eq:Complexity2}
\mathrm{SCM} = \mathrm{WPE} \cdot Q_J(P, P_e),
\end{equation} 

where WPE represents the normalized Weighted Permutation Entropy (Eq.~\ref{eq:W_Permutation_Entropy}) and $Q_J(P, P_e)$ is the disequilibrium, defined in terms of the Jensen-Shannon divergence:  

\begin{equation}
\label{eq:Q-disequilibrium}
Q_J(P,P_e) = Q_0 \cdot J(P,P_e),
\end{equation}

with  

\begin{equation}
\label{eq:Jensen-Shannon-Divergence}
J(P, P_e) = S\left(\frac{P+P_e}{2}\right) - \frac{S(P)}{2} - \frac{S(P_e)}{2}.
\end{equation}

Here, $Q_0$ is a normalization constant ($0 \leq Q_J \leq 1$), given by the inverse of the maximum possible value of $J(P, P_e)$, i.e., $Q_0 = 1 / J(P_0, P_e)$. The function $S$ represents the Shannon Entropy (Eq.~\ref{eq:ShannonEntropy1}), where $P_e$ denotes the equiprobable probability distribution (all symbols have the same probability of occurrence), 
$P_0$ corresponds to a Dirac delta distribution, where only one element has a nonzero probability, and $P$ is the empirical probability distribution constructed according to Eq.~\ref{eq:PDF_WPE}.

The statistical complexity measure offers complementary insights to entropy, as it compares our probability density function $P$ with an equiprobable one ($P_e$). Additionally, it can be shown that for a fixed value of normalized entropy, the associated complexity lies within a range bounded by $C_{\mathrm{min}}$ and $C_{\mathrm{max}}$, whose limits are determined solely by the embedding dimension (D)~\cite{RossoPRL, lamberti2004intensive, martin2006generalized}.

\section{The Data}
\subsection{Experimental Time Series}
\label{Sec:Experimental}

\begin{figure}
  \centering
  \begin{minipage}{0.25\linewidth}
    \begin{flushleft}(a)\end{flushleft}
    \includegraphics[width=\linewidth]{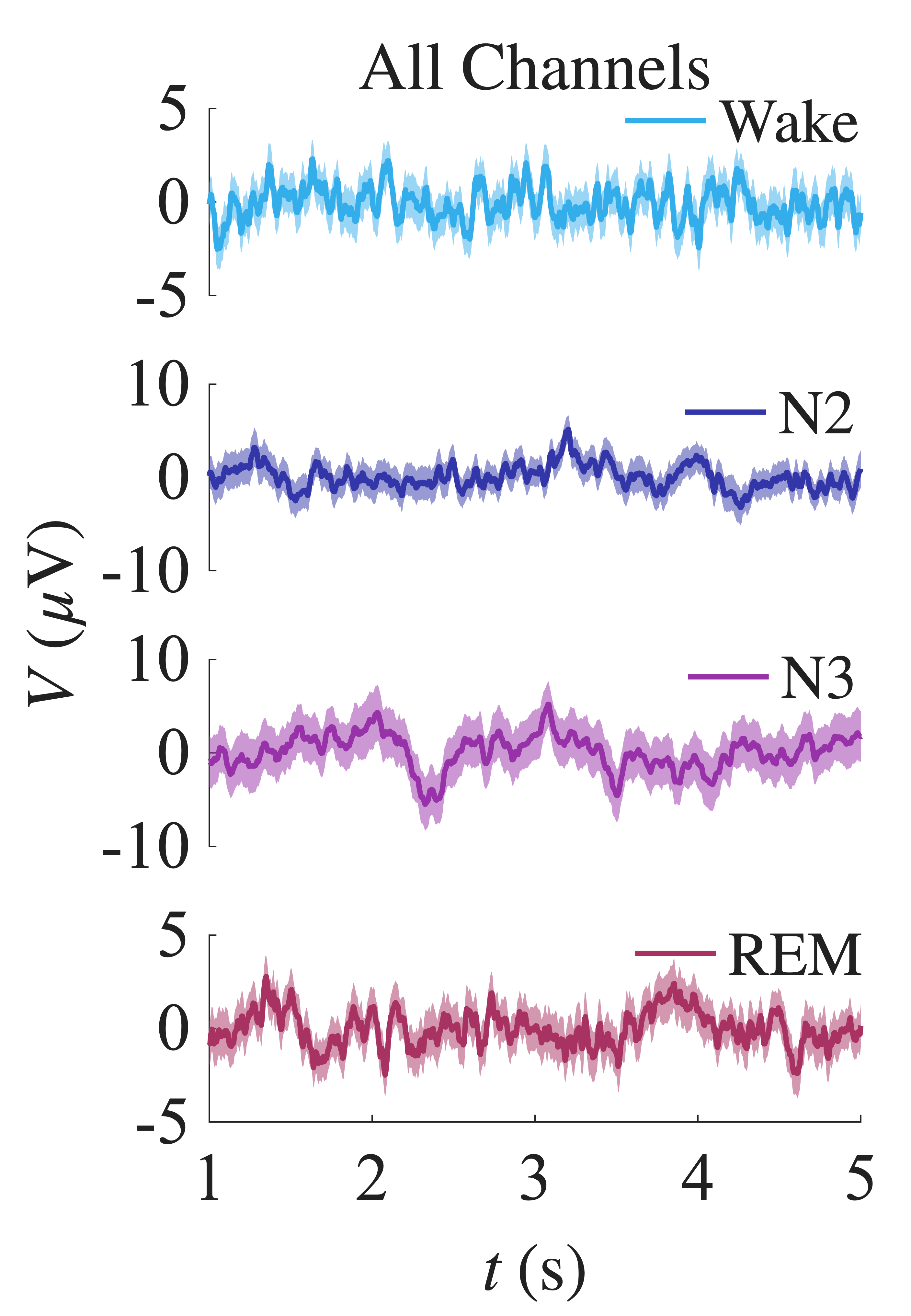}
  \end{minipage}
  \begin{minipage}{0.25\linewidth}
    \begin{flushleft}(b)\end{flushleft}
    \includegraphics[width=\linewidth]{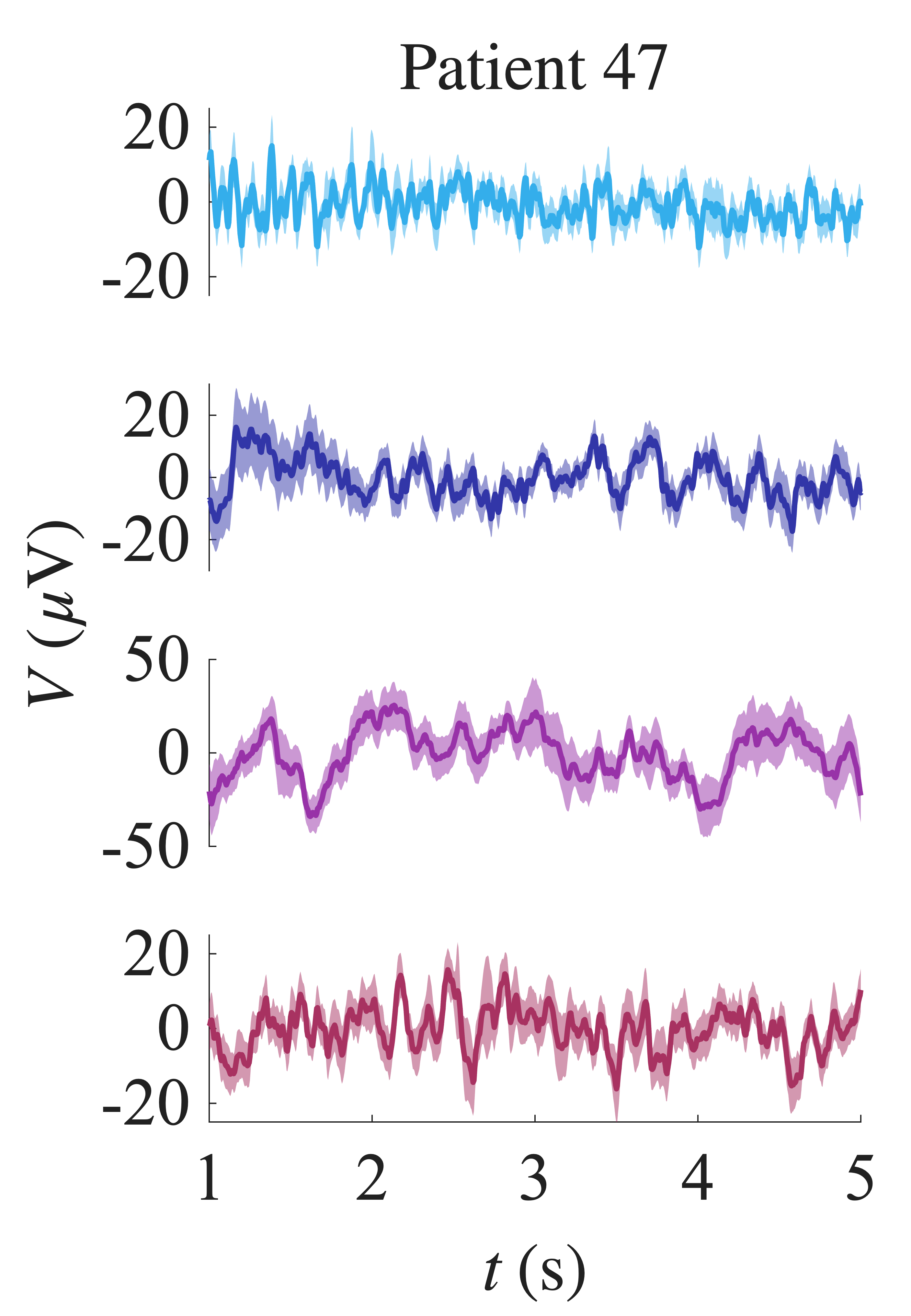}
  \end{minipage}
  \begin{minipage}{0.25\linewidth}
    \begin{flushleft}(c)\end{flushleft}
    \includegraphics[width=\linewidth]{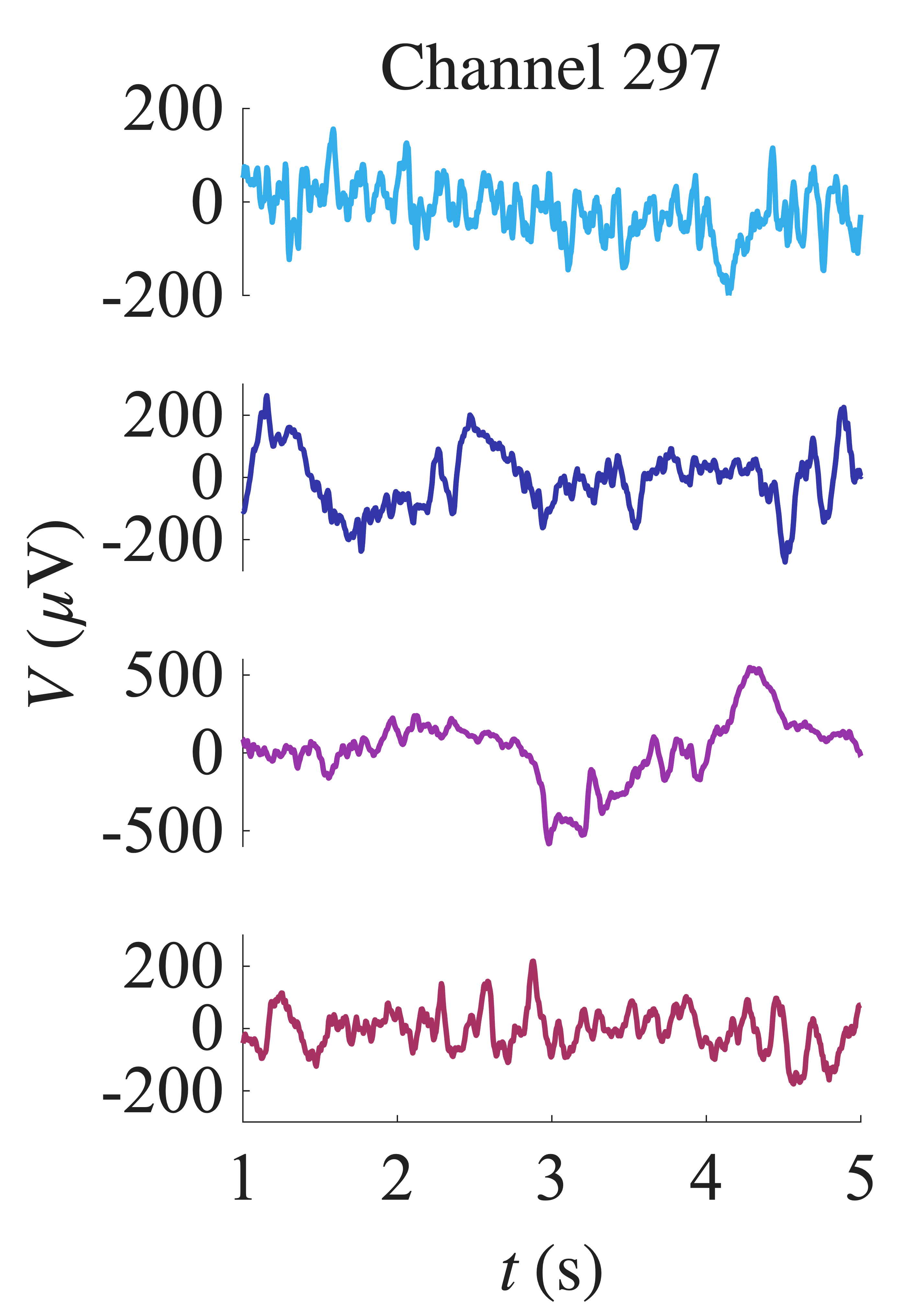}
  \end{minipage}
\caption{Representative time series of sleep stages. (a) Five-second segment of the average signal across all channels, including the mean (solid line) and standard deviation (shaded area) for each of the four sleep stages. (b) Five-second segment showing the mean (solid line) and standard deviation (shaded area) of the electrical signals across all channels from a single participant (patient 47), also for the four studied stages. (c) Five-second segment of the signal from an individual channel (channel 297) across the same sleep stages, illustrating the similarity between the average and individual dynamics.}
  \label{fig:time_series}
\end{figure}

\begin{figure}
  \centering   \includegraphics[width=0.5\linewidth]{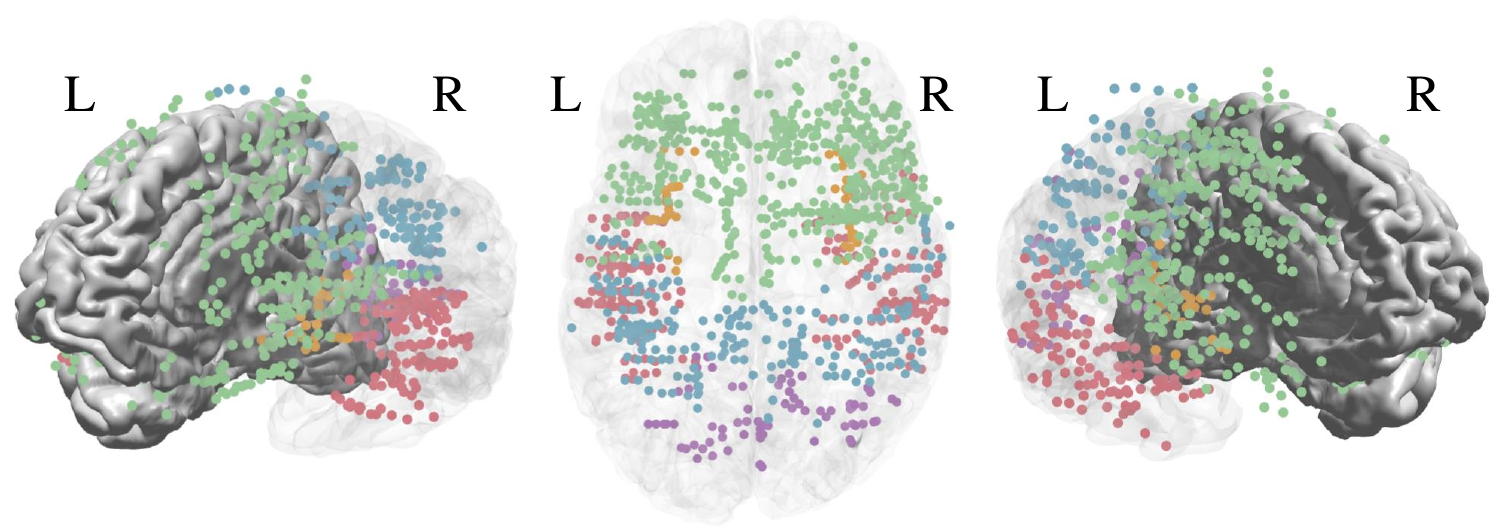}  
  \caption{Location of the 1,012 intracranial channels. The colors represent different brain lobes: occipital (light purple), parietal (light blue), insula (light orange), frontal (light green), and temporal (light red).}
  \label{fig:position_channels}
\end{figure}

The experimental data used in this work were obtained from a public and online database, available at:  \href{https://mni-open-ieegatlas.research.mcgill.ca/main.php?}{MNI Open iEEG Atlas}. The project results from a collaboration between the Montreal Neurological Institute and Hospital, the Grenoble-Alpes University Hospital, and the Centre hospitalier de l'Université de Montréal. The database contains intracranial electroencephalography (iEEG) recordings from 106 patients with drug-resistant focal epilepsy, that is, patients who did not respond adequately to conventional clinical treatments and who, for this reason, were referred for surgical evaluation, which enabled intracranial monitoring~\cite{Frauscher2018, frauscher2018high, von2020human}. The sample includes individuals aged between 13 and 62 years and includes individuals of both sexes. The recordings were obtained with intracerebral depth electrodes (Dixi, MNI, or AdTech) and with subdural strip and grid electrodes (AdTech), according to the clinical needs of each patient~\cite{Frauscher2018, frauscher2018high, von2020human}. Ethical approval was obtained from the Montreal Neurological Institute, and the acquisitions were performed after a minimum interval of 72 hours following electrode implantation, or one week in the case of other devices, minimizing residual effects of anesthesia or the surgical procedure. The time series correspond to four brain states: wakefulness (Wake), non-REM sleep (stages N2 and N3), and REM sleep. Stage N1 was not considered because it is not part of the set of stages made available in the MNI Open iEEG Atlas.

Only channels located in regions not affected by the pathology were analyzed, totaling up to 1772 channels in the wakefulness condition, 1468 in stages N2 and N3, and 1012 in the REM stage. The selection of these channels followed rigorous criteria~\cite{Frauscher2018}: (i) each channel had to be located in structurally normal tissue on magnetic resonance imaging, that is, outside any visible lesional area and outside the clinically defined seizure onset zone; (ii) the channel could not exhibit interictal discharges such as spikes, sharp waves, or spike-wave complexes at any point during the analyzed recording, which means that the entire trace, including periods of wakefulness and sleep, was inspected; and (iii) there should be no abnormal slowing or other electrophysiological alterations suggestive of cortical dysfunction. Nevertheless, an inherent limitation of the database should be explicitly stated: the participants are not healthy volunteers. Thus, the term "normal" refers to probably normal activity, defined based on the clinical, anatomical, and electrophysiological criteria cited above~\cite{Frauscher2018}.

Signal preprocessing involved resampling to 200 Hz, with an 80 Hz anti-aliasing low-pass filter, and line noise was reduced by adaptive filtering~\cite{Frauscher2018, frauscher2018high, von2020human}. Artifact inspection and removal were performed visually by a neurophysiologist, and the artifact-free segments were concatenated with 2-s zero-amplitude buffers. All channels were then padded with zeros to a fixed length of 68 s (13,600 samples)~\cite{Frauscher2018, frauscher2018high, von2020human}. Since these zero-valued samples do not correspond to recorded activity, they were \novo{treated as missing data and excluded from all analyses. Buffers and padding were identified as runs of at least 10 consecutive zero-valued samples (50 ms), so that each series consists of one or more contiguous segments of recorded activity. Ordinal patterns were extracted only within these segments, and their (weighted) counts were pooled across segments before normalization, so that no ordinal vector spans samples that were not contiguous in the original recording.}

The electrode positions were transformed into the same anatomical reference space, the symmetric MNI ICBM 2009a template, with a resolution of 1×1×1 mm, and the identification of anatomical regions was performed automatically based on the MICCAI 2012 atlas. At the end of this process, each channel had standardized three-dimensional coordinates (x, y, z) in MNI space, as well as an anatomical label corresponding to one of the 38 cortical gray matter regions defined in the atlas~\cite{Frauscher2018, frauscher2018high, von2020human}. For analysis and visualization purposes, we grouped these 38 regions into five macro-regions, corresponding to the four main cerebral lobes and the insula. The distribution of channels by lobe is: frontal (469 channels), parietal (204), temporal (221), occipital (69), and insula (49). The spatial locations of these channels are shown in Figure~\ref{fig:position_channels}.

For this work, only channels with data simultaneously available in all four sleep stages were included, which reduced the dataset from up to 1772 to 1012 channels and the number of patients from 106 to 65. The number of valid channels is not uniform among patients, ranging from a single channel to 43 channels, with patient 47 having the greatest coverage. \novo{Because entropy and complexity estimates depend on series length, every series was truncated to its first 10,800 valid samples (54 s at 200 Hz), the minimum amount of recorded activity available across all channels and stages. In averaged series, each time point was computed over the channels with recorded activity at that instant, and a new segment was started whenever the set of contributing channels changed. The number of ordinal vectors therefore varies across series and decreases with the embedding delay $\tau$ (see Supplementary Note S1).}

In summary, 1012 channels were analyzed, distributed among 65 patients and five macro-regions, in each of the four sleep stages, totaling 4048 time series\novo{, each containing 10,800 valid samples}. Fig.~\ref{fig:time_series}(a) presents a five-second segment of the time series, with the mean and standard deviation across all channels for the four stages. To illustrate the correspondence between the mean signal and individual behavior, Figs.~\ref{fig:time_series}(b) and (c) present five-second segments from the representative participant (patient 47) and from a representative individual channel (channel 297), respectively. From this point onward, the color scheme used to represent each stage follows the same pattern as in Fig.~\ref{fig:time_series}.

\subsection{Computational modeling}
\label{Sec:simulation}

To simulate whole-brain dynamics, we modeled each cortical region as a node using a mean-field (MF) description of a network of Adaptive Exponential Integrate and Fire (AdEx) neurons~\cite{Brette05,el2009master,carlu2020mean}. The comparison with empirical iEEG data is therefore made at the level of the dynamical organization of the time series, rather than as a direct reproduction of the iEEG signal. The MF framework captures the evolution of the firing rate of interacting local excitatory (e) and inhibitory (i) populations, including activity-dependent neuronal adaptation, as follows:

\begin{equation*} 
\label{eq:MF}
    \begin{split}
        T\frac{\partial\nu_\mu}{\partial t} &= (F_\mu-\nu_\mu) + \frac{1}{2}c_{\lambda\eta}\frac{\partial^2F_\mu}{\partial\lambda\partial\eta}, \\
        T\frac{\partial c_{\lambda\eta}}{\partial t} &= \frac{F_\lambda(T^{-1}-F_\eta)}{N_{\lambda}}+(F_\lambda-\nu_\lambda)(F_\eta-\nu_\eta)\\
       &+\frac{\partial F_\lambda} {\partial\nu_\mu}c_{\eta\mu}+\frac{\partial F_\eta} {\partial\nu_\mu}c_{\lambda\mu}-2c_{\lambda\eta},\\
        \frac{\partial W_\mu}{\partial t} &= -\frac{W_\mu}{\tau_{W}}+b_\mu\nu_\mu+\frac{a_\mu\left[\mu_V(\eta_e,\eta_i,W_\mu)-E_{L\mu}\right]}{\tau_{W}},
    \end{split}
\end{equation*}

where $\nu_\mu$ is the mean firing rate of population $\mu=\{e,i\}$, $c_{\lambda\eta}$ is the covariance between population $\lambda$ and $\eta$, $W_\mu$ is the mean adaptation, $b_\mu$ is the adaptation level, $a_\mu$ is the subthrehsold adaptation,  $T$ is the MF characteristic time constant. $F_\mu=F_\mu(\nu_e,\nu_i,W_\mu)$ is the transfer function (TF) of a given neuron type $\mu$. Excitatory and inhibitory neurons were modeled as regular-spiking (RS) and fast-spiking (FS), respectively~\cite{di2019biologically}. The TF characterizes the dependence of the output firing rate on the excitatory ($\nu_e$) and inhibitory ($\nu_i$) inputs, and can be written as a function of its mean subthreshold membrane voltage $\mu_V$, its standard deviation $\sigma_V$, and its time correlation time decay $\tau_V$\cite{zerlaut2018modeling}:

\begin{equation*}
\label{eq:TF}
    F = \nu_{\mathrm{out}} = \frac{1}{2\tau_V}\cdot \operatorname{erfc}\left(\frac{V_{\mathrm{thr}}^{\mathrm{eff}}-\mu_V}{\sqrt{2}\sigma_V}\right),
\end{equation*}
where ($\mu_V$,$\sigma_V$,$\tau_V$) are obtained by solving a set of equations as described in~\cite{di2019biologically}. $V_{\mathrm{thr}}^{\mathrm{eff}}$ is a phenomenological spike threshold voltage taken as a second-order polynomial:

\begin{equation*}
\begin{aligned}
V_{\mathrm{thr}}^{\mathrm{eff}}(\mu_V,\sigma_V,\tau_V^N)
&= P_0
 + \sum_{x\in\{\mu_V,\sigma_V,\tau_V^N\}}
   P_x\left(\frac{x-x_0}{\delta x_0}\right) \\
&\quad
 + \sum_{x,y\in\{\mu_V,\sigma_V,\tau_V^N\}^2}
   P_{xy}
   \left(\frac{x-x_0}{\delta x_0}\right)
   \left(\frac{y-y_0}{\delta y_0}\right).
\end{aligned}
\end{equation*}

with $\tau_V^N=\frac{\tau_Vg_l}{C_m}$, where $g_l$ is the leakage conductance and $C_m$ the membrane capacitance. The constant values used were the same obtained in~\cite{di2019biologically}: $\mu_V^0=-60$~mV, $\sigma_V^0=0.004$~mV, $(\tau^N_V)^0=0.5$, $\delta\mu^0_V=0.001$~mV,
$\delta\sigma^0_V=0.006$~mV, and $\delta(\sigma^N_V)^0=1$. Accordingly, the fitted polynomials $P$, for the excitatory (e) and inhibitory (i) types of neurons are: $P_0^{e,i}=(-0.0498,-0.0514)$, $P_{\mu_V}^{e,i}= (0.00506,0.004)$, $P_{\sigma_V}^{e,i}=(-0.025,-0.0083)$, $P_{\tau_V}^{e,i}=(0.0014,0.0002)$, $P_{\mu^2_V}^{e,i}=(-0.00041,-0.0005)$, $P_{\sigma^2_V}^{e,i}=(0.0105,0.0014)$, $P_{\tau^2_V}^{e,i}=(-0.036,-0.0146)$, $P_{\mu_V\sigma_V}=(0.0074,0.0045)$, $P_{\mu_V\tau_V}=(0.0012,0.0028)$, $P_{\sigma_V\tau_V}=(-0.0407,-0.0153)$. $\mu_V$ is a function that represents the average membrane potential of a given population:

\begin{align*}
    \mu_V=\frac{\mu_{Ge}E_e+\mu_{Gi}E_i+g_LE_L-W}{\mu_{Ge}+\mu_{Gi}+g_L},
\end{align*}

where $\mu_{Ge}=\nu_eK_eu_eQ_e$, and similarly to the population $i$. $Q_\mu$ is the conductance weight, $u_\mu$ is the synaptic time decay, and $K_\mu=Np$ is a constant that depends on the number $N=10^4$ of neurons and the probability $p=0.05$ of connection. All parameters were obtained from~\cite{di2019biologically,sacha2025computational} and are summarized in Table~\ref{table:paramMF}. 

Previous equations describe the population dynamics of a single cortical region of excitatory and inhibitory populations. To extend it to account for a large network of interconnected cortical regions, the network transfer function can be rewritten as $F_\mu(\nu_e^{\mathrm{in}}(k),\nu_i(k),W(k))$:

\begin{equation*}
\begin{aligned}
\nu^{\mathrm{in}}_\mu(k,t)
&= \nu_e(k,t)+\nu_{\mathrm{aff}}(k,t) \\
&\quad + G\sum_j C_{kj}\nu_e\!\left(j,t-D_{kj}\right).
\end{aligned}
\end{equation*}

where the sum runs over all nodes $j$, $C_{kj}$ is the connection strength between nodes $j$ and $k$, and $D_{kj}$ is the matrix of delays. $G$ is a constant coupling factor that rescales the connection strength while maintaining the ratio between different values. $\nu_{\mathrm{aff}} = \nu_{\mathrm{drive}}+\sigma \zeta(t)$, represent the afferent input, where $\nu_{\mathrm{drive}}$ is constant external input, $\sigma = 4$ is the noise weight, and $\zeta$ denotes an Ornstein-Uhlenbeck process: $\zeta(t)=-(\zeta(t)dt/\tau_{\mathrm{OU}})+dW_t$, with $dW_t$ representing a Wiener process of amplitude one and average zero.

The connectivity between each MF node was defined by human tractography data from the Berlin empirical data, and the Desikan-Killiany parcellation was used to define $68$ cortical regions~\cite{desikan2006automated}. Connections among cortical regions and propagation delays were defined by tract lengths and with estimates in human diffusion tensor imaging (DTI) data~\cite{sanz2015mathematical,schirner2015automated}.

This MF model constrained by the human connectome has been shown to replicate the dynamics of different brain states, such as awake- and sleep-like states~\cite{goldman2023simulation,sacha2025computational,DallaPorta2025WBM}. Both states are dependent on the level of adaptation ($b$), such that for low values the dynamics are more asynchronous, standing for awake-like states. For high levels of $b$ the system displays slow wave dynamics, a fingerprint of slow wave sleep (NREM phase $3$).

Here, $b$ should be regarded as a control parameter spanning a wake-like to
sleep-like dynamical continuum, rather than as a direct representation of
individual sleep stages. To simulate a continuous transition among these two
states, we varied $b$ from 40 to 95~pA in steps of 5~pA.
\novo{The model was integrated with a time step of 1~ms for a total of
60~s. For each value of $b$, we ran five independent noise realizations, keeping all other parameters
fixed.} For each value of $b$, we analyzed the mean firing-rate time series of
the excitatory population, $\nu_e(t)$, from each of the 68 cortical regions.
The simulated signals were resampled to 200~Hz, \novo{and
analyzed using segments of 10{,}800 samples (54~s), matching the effective
length of the empirical series (Sec.~\ref{Sec:Experimental})}. All the simulations were performed using The Virtual Brain software~\cite{sanz2015mathematical,schirner2022brain}.

\begin{table}[!h]
\caption{AdEx mean-field parameters}
\centering
\begin{adjustbox}{max width=\linewidth}
\begin{tabular}{|cccc|}
\hline
Parameter & Value & Parameter & Value \\
\hline
$T$ & $20$~ms & $C_m$ & $200$~pF\\
$E_{L;e,i}$ & $\{-64,-65\}$~mV & $Q_{e,i}$ & $\{1.5,5\}$~nS\\
$N_{e,i}$ & $\{8,2\}\times10^3$ & $\tau_{e,i}$ & $\{5,5\}$~ms\\
$p$ & $5\%$ & $\nu_{e,i}^{\mathrm{ext}}$ & $\{0.315,0.315\}$~Hz\\
$b_{e}$ & varied & $K_{e,i}$ & $\{400,0\}$\\
$b_i$ & $0$~pA & $E_{e,i}$ & $\{0,-80\}$\\
$a_{e,i}$ & $\{0,0\}$~nS & $\tau_{\mathrm{OU}}$ & $5$~ms\\
$\tau_w$ & $500$~ms & noise weight & $3\times10^{-4}$~nS\\
$g_L$ & $10$~nS & conduction speed & $4$~m/s\\
\hline
\end{tabular}
\end{adjustbox}
\label{table:paramMF}
\end{table}

\section{Results}
\label{Sec:Results}

\subsection{Channel-level analysis}

\begin{figure}
  \centering
  \begin{minipage}[t]{0.49\linewidth}
    \begin{flushleft}(a)\end{flushleft}
    \includegraphics[width=\linewidth]{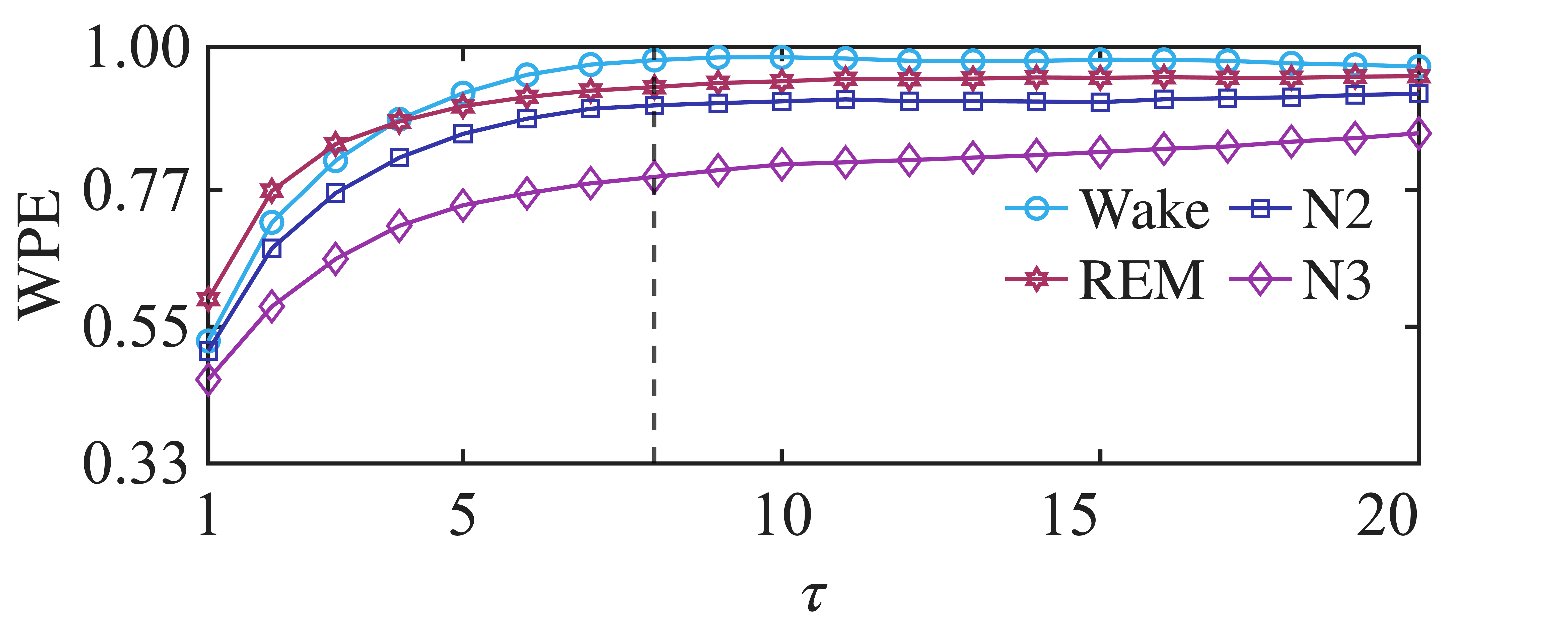}
  \end{minipage}
  \begin{minipage}[t]{0.49\linewidth}
    \begin{flushleft}(b)\end{flushleft}
    \includegraphics[width=\linewidth]{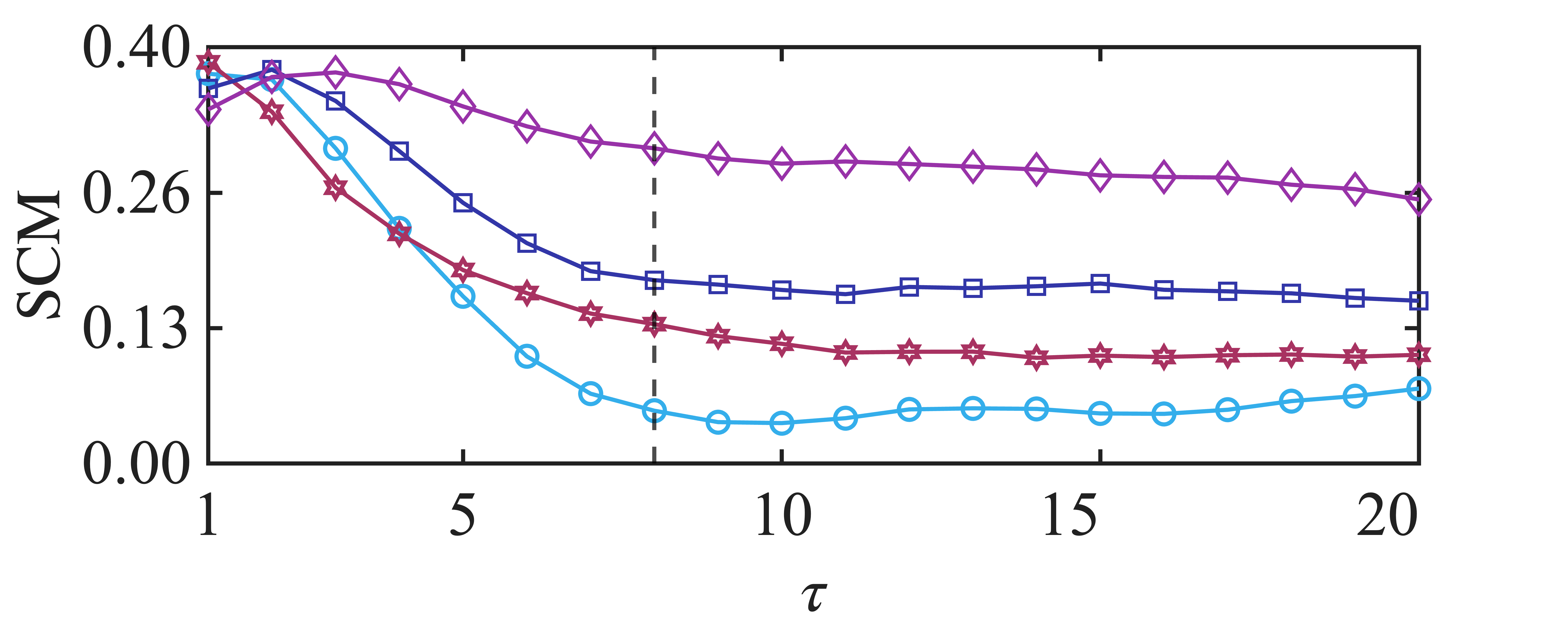}
  \end{minipage}
\caption{Variation of (a) Weighted Permutation Entropy (WPE) and 
  (b) Statistical Complexity Measure (SCM) as a function of the embedding 
  delay $\tau$. The measures were computed from the mean iEEG signal 
  obtained by averaging across the 1,012 recording channels for each brain 
  state. Values of $\tau$ range from 1 to \novo{20}, corresponding to delays from 
  5 to \novo{100}~ms at a sampling frequency of 200~Hz. The four curves correspond 
  to Wake, REM, N2, and N3. The dashed vertical line indicates $\tau=\novo{8}$ 
  (\novo{40}~ms), which maximized the overall separation among brain states \novo{(see Supplementary Fig.~S1(b))}.}
  \label{fig:PE_SCM_exp}
\end{figure}

For all results presented below, we adopted an embedding dimension of $D=6$, as it provides a good balance between the richness of the captured temporal patterns and adequate statistical sampling of the $D!$ possible permutations. \novo{The main patient-level and regional results were also reproduced with $D = 4$ and $D = 5$ (Supplementary Note S2).}

As a first approach, WPE and SCM were computed from the average signal of the $1,012$ channels. In other words, each of the 1,012 channels is represented by its corresponding time series, resulting in a total of 1,012 time series. These signals were then combined by averaging them, yielding a single representative time series for each sleep stage (Fig.~\ref{fig:time_series} (a)). The WPE and SCM measures were subsequently calculated from this averaged signal. 

To determine an appropriate temporal scale for the ordinal analysis, we systematically varied the embedding delay $\tau$ from 1 to 100, corresponding to 5--500~ms at a sampling frequency of 200~Hz. Figure~\ref{fig:PE_SCM_exp} shows the resulting (a) WPE and (b) SCM values \novo{for $\tau \le 20$ (5--100~ms)}. \novo{Because ordinal vectors spanning a buffer are excluded, the number of vectors available decreases with $\tau$; for $\tau \ge 18$ the Wake series has fewer than $10\cdot D!$ vectors (Supplementary Note S1), so the behavior of the curves at larger delays should not be interpreted.} Both measures exhibit a clear dependence on the embedding delay, with enhanced separation among brain states around $\tau = \novo{8}$, indicated by the dashed vertical line.

To quantify this separation, for each value of $\tau$ we calculated the Euclidean distances between all six pairs of brain states (Wake, REM, N2, and N3) points in the joint SCM--WPE plane and used the sum of these distances as an overall measure of state discrimination. \novo{Among the delays for which all stages retain at least $10\cdot D!$ ordinal vectors ($\tau \le 17$), this quantity reached its maximum at $\tau = 8$, corresponding to a 40-ms delay (Supplementary Note S1, Fig.~S1(b)).} We therefore, unless otherwise \novo{stated}, selected $\tau = \novo{8}$ as a data-driven embedding delay for subsequent analyses. \novo{Because this delay was selected from the same data used in the subsequent analyses, we verified that the main conclusions hold over a broad range of delays and embedding dimensions (Supplementary Note S2)}.

\begin{figure}
    \centering
    \includegraphics[width=1\linewidth]{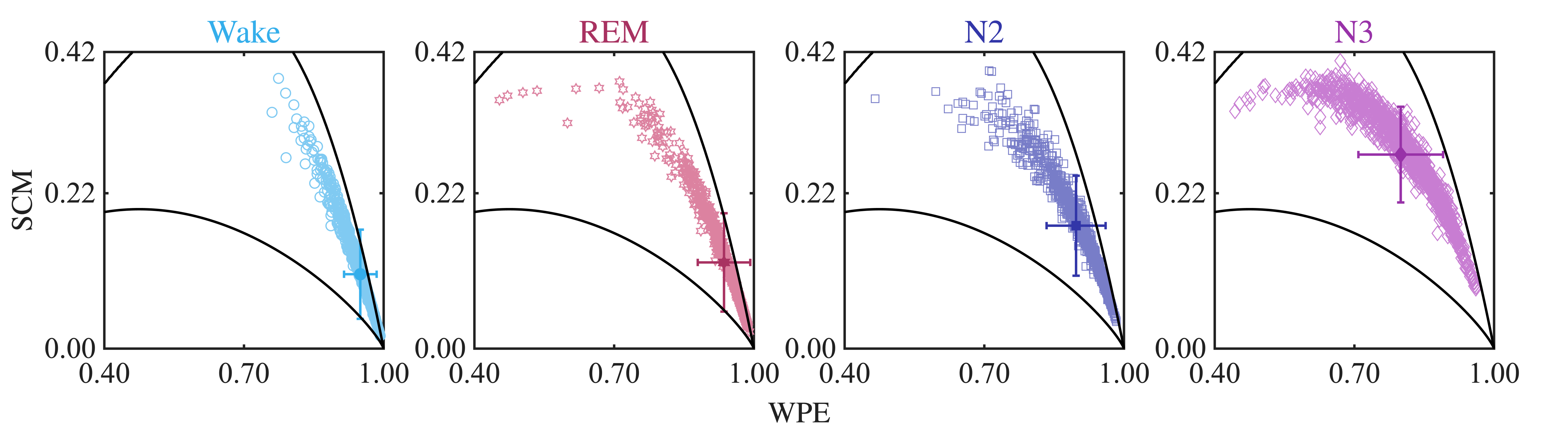}
    \caption{Distribution of the 1,012 channels in the complexity--entropy plane for $\tau=\novo{8}$ and $D=6$, in Wake, REM, N2, and N3. The semitransparent points represent the WPE and SCM values of each channel, whereas the opaque points and bars indicate the mean and standard deviation across channels, the same as those presented later in Fig.~\ref{fig:plano_exp}(b). The black curves correspond to the theoretical limits $C_{\max}$ and $C_{\min}$.}
    \label{fig:allchannels_subplot_juntos}
\end{figure}

Before any averaging \novo{with $\tau = 8$} procedure, Fig.~\ref{fig:allchannels_subplot_juntos} shows the position of each of the 1,012 channels in the SCM$\times$WPE plane at each stage. Three features stand out: (i) in all stages, the channels are not freely distributed throughout the plane, but are concentrated in a narrow band close to the upper limit $C_{\max}$, so that differences between states are manifested mainly as displacements along this band. (ii) despite the variability across channels, the point cloud shifts progressively from Wake to N3 toward lower WPE and higher SCM. In Wake, the vast majority of channels occupy the high-entropy and low-complexity end. In REM, the distribution is similar, although some channels extend to WPE values close to 0.5. And finally, (iii) the dispersion increases as sleep deepens, that is, in N2 the cloud elongates along the band and, in N3, occupies practically its entire extent, with WPE ranging from values close to 1 down to about 0.4. This greater heterogeneity during NREM sleep is consistent with the local nature of slow waves, whose amplitude, synchronization, and recruitment vary across cortical regions~\cite{Nir2011NREM, bernardi2018local, siclari2017local}, an aspect revisited in greater detail in the regional analysis.

\begin{figure}
 \centering
  \begin{minipage}[t]{0.4\linewidth}
  \begin{flushleft}(a)\end{flushleft}
  \includegraphics[width=\linewidth]{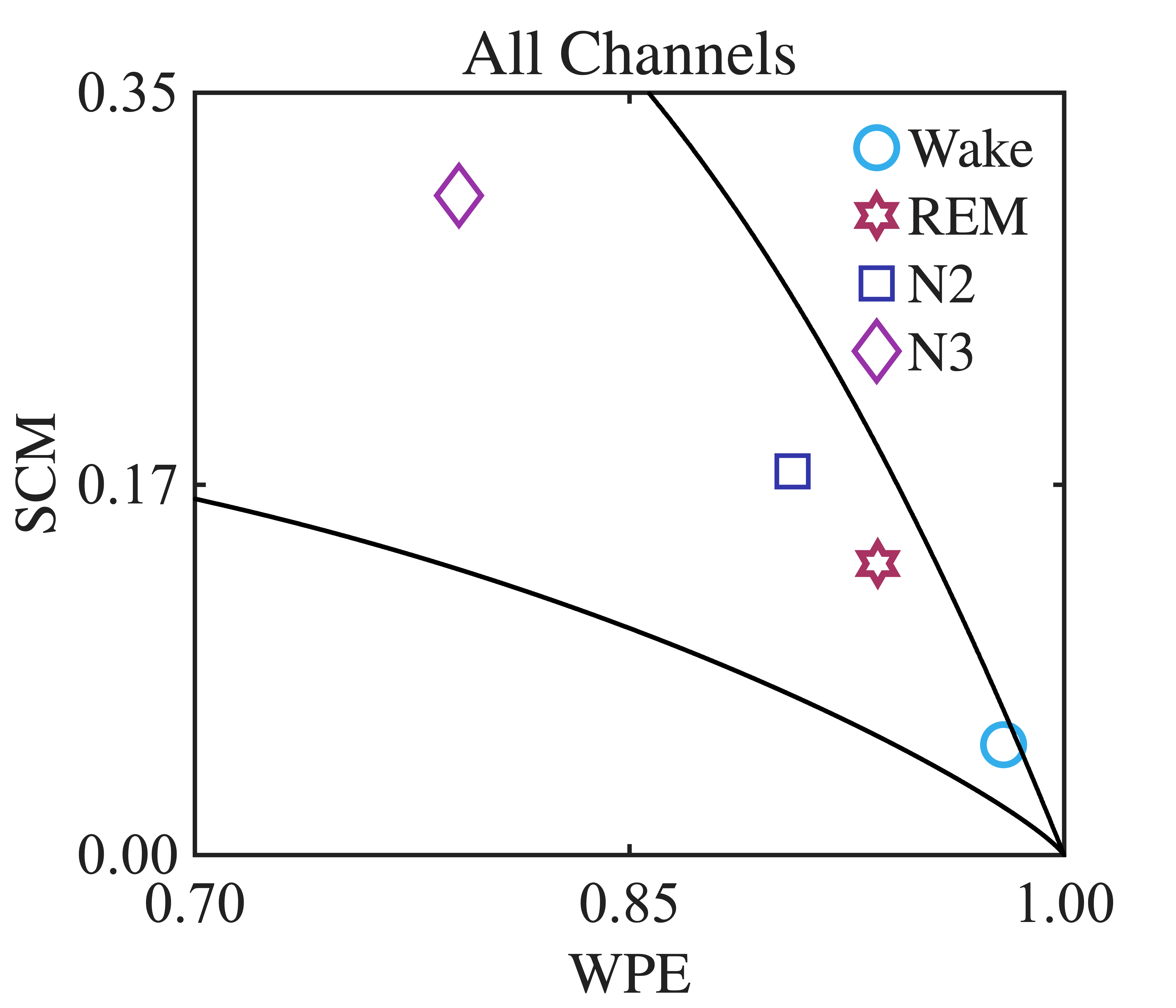}
  \end{minipage}
  \centering
  \begin{minipage}[t]{0.4\linewidth}
  \begin{flushleft}(b)\end{flushleft}
  \includegraphics[width=\linewidth]{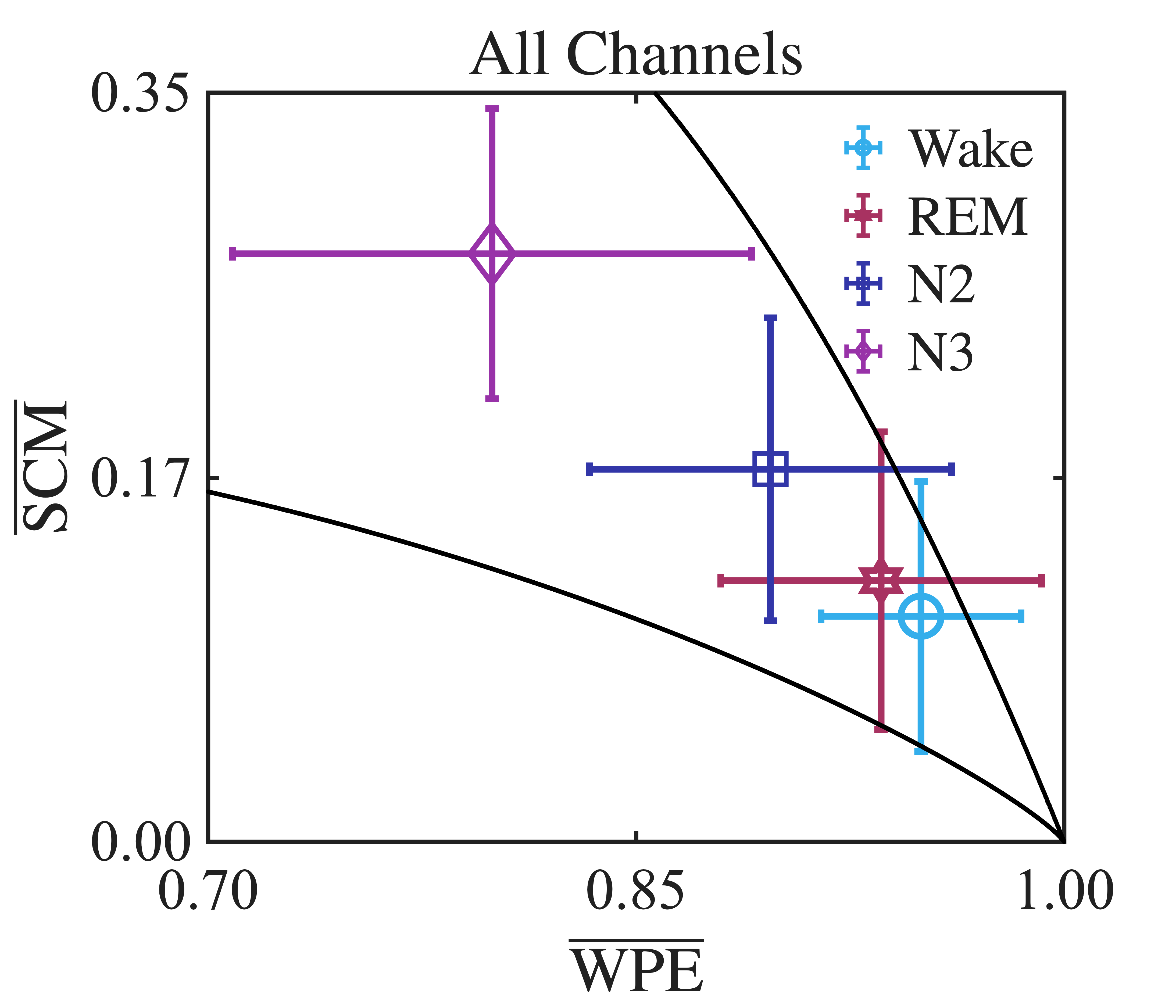}
  \end{minipage}  
\caption{Complexity--entropy plane for the four sleep stages (Wake, REM, N2, and N3), obtained using an embedding delay of $\tau=\novo{8}$ and $D = 6$. (a) WPE and SCM computed from the mean iEEG signal obtained by averaging across all 1,012 channels for each sleep stage. (b) WPE and SCM computed separately for each channel and subsequently averaged across channels; error bars represent the variability across channels. The black curves indicate the theoretical minimum and maximum bounds of the statistical complexity for the adopted embedding dimension.}
\label{fig:plano_exp}
\end{figure}

After selecting $\tau=\novo{8}$ as the temporal scale for the subsequent analyses, we projected the WPE and SCM values onto the complexity--entropy plane for the different sleep stages. Two distinct averaging strategies were considered. In the first, shown in Figure~\ref{fig:plano_exp}(a), iEEG from all channels were first averaged for each sleep stage, and WPE and SCM were then computed from the resulting time series. In the second approach, shown in Figure~\ref{fig:plano_exp}(b), WPE and SCM were computed separately for each channel and sleep stage, after which the metric values were averaged across channels, yielding a mean pair $\overline{\mathrm{WPE}}$ and $\overline{\mathrm{SCM}}$. 

Despite the differences between these two strategies, both revealed the same qualitative organization of the states in the complexity--entropy plane. A systematic progression was observed along the sequence Wake $\rightarrow$ REM $\rightarrow$ N2 $\rightarrow$ N3, characterized by a decrease in WPE accompanied by an increase in SCM. The Wake state occupies the region with the highest WPE values and the lowest SCM values, whereas N3 lies at the opposite end, with lower WPE and higher statistical complexity. REM and N2 occupy intermediate positions, thereby establishing a continuous trajectory between these two regimes.

Using our WPE and SCM metrics jointly, this trajectory suggests a progressive reorganization of cortical temporal dynamics across sleep. In our analysis, higher WPE values indicate that a broader set of ordinal patterns contributes to the signal dynamics, making its temporal evolution less predictable. Our SCM metric, in turn, does not simply quantify signal regularity, but depends jointly on entropy and on the deviation of the pattern distribution from equiprobability. Therefore, the increase in SCM observed as WPE decreases indicates, within the framework of our metrics, that reduced unpredictability is accompanied by the emergence of a more structured and non-uniform distribution of temporal patterns, rather than merely by a nonspecific loss of variability.

The organization of the states in the complexity--entropy plane is consistent with known changes in cortical dynamics across the sleep--wake cycle. Wakefulness, characterized by high WPE and low SCM, exhibits a greater diversity of temporal patterns, consistent with the relatively fast and desynchronized cortical activity characteristic of this state~\cite{brown2012control,harris2011cortical,sanchezvives2025multiscale}. REM sleep remains close to wakefulness, reflecting its similarly fast and desynchronized cortical activity, although the shift toward lower WPE and higher SCM indicates a distinct temporal organization~\cite{brown2012control,nir2010dreaming}.

The progression toward NREM sleep is accompanied by lower WPE and higher SCM. In N2, this behavior is consistent with the recurrent presence of characteristic temporal structures, such as sleep spindles and K-complexes~\cite{de2003sleep,cash2009human}. This trend becomes even more pronounced in N3, where the predominance of slow oscillations and the increased synchronization of neuronal populations give rise to more regular and structured temporal dynamics~\cite{massimini2004sleep,brown2012control}. Thus, the increase in SCM along the transition toward deep sleep should be understood as reflecting greater statistical structure in the distribution of ordinal patterns, rather than a general increase in the ``complexity'' of brain activity.

In Figure~\ref{fig:plano_exp}(b), a considerable dispersion of WPE and SCM values across channels can also be observed, particularly for stages N2 and N3, along with a partial overlap between the error bars of some states, especially Wake and REM. This variability is expected, since the recordings were obtained from different participants and cortical regions and therefore incorporate spatial and interindividual heterogeneity in brain dynamics. Despite this dispersion, the mean values preserve the same ordering observed when the metrics are calculated from the averaged signal. Thus, both analysis strategies point to a consistent global trend: the progression from wakefulness to deep NREM sleep is accompanied by a reduction in the diversity of temporal patterns and, simultaneously, by an increase in their statistical organization.

\subsection{Patient-level analysis}

Interpreting the 1,012 channels as independent observations requires caution because they are distributed across only 65 patients. Therefore, several channels belong to the same individual and share anatomical, physiological, and experimental characteristics. In addition, the number of valid channels varies across patients, meaning that patients with more electrodes would have greater influence on the analysis. For this reason, it is important to account for the hierarchical structure of the dataset and to assess whether the observed pattern remains when each patient is treated as a single unit of observation. To address these issues, we treated each of the 65 patients as an independent unit of observation and evaluated, for each patient individually, the trajectory of sleep stages in terms of WPE and SCM. Our goal was to determine whether the global pattern Wake $\rightarrow$ REM $\rightarrow$ N2 $\rightarrow$ N3 remained preserved at the individual level.

\begin{figure}
\centering
\begin{minipage}[t]{0.36\linewidth}
\begin{flushleft}(a)\end{flushleft}
\includegraphics[width=\linewidth]{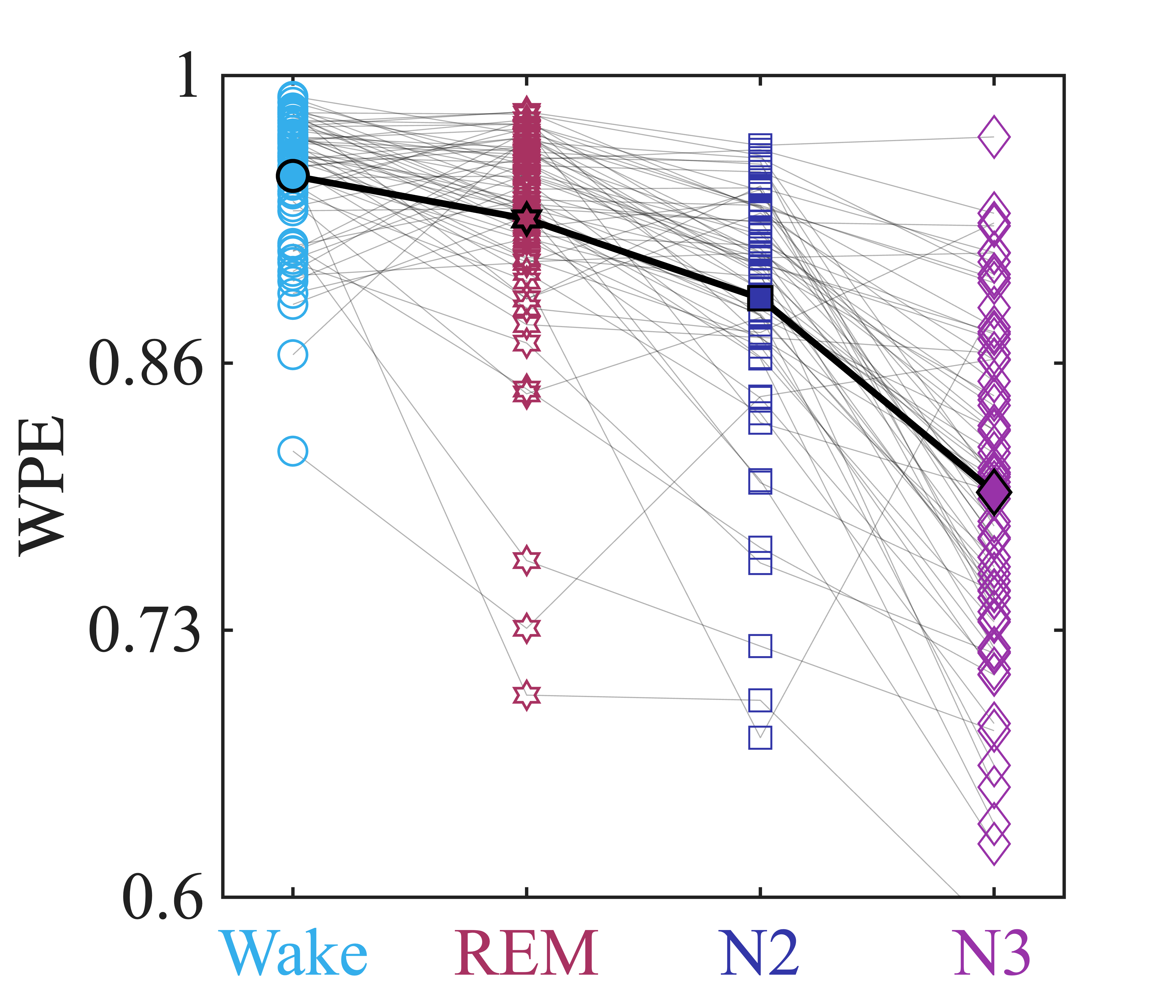}
\end{minipage}
\begin{minipage}[t]{0.36\linewidth}
\begin{flushleft}(b)\end{flushleft}
\includegraphics[width=\linewidth]{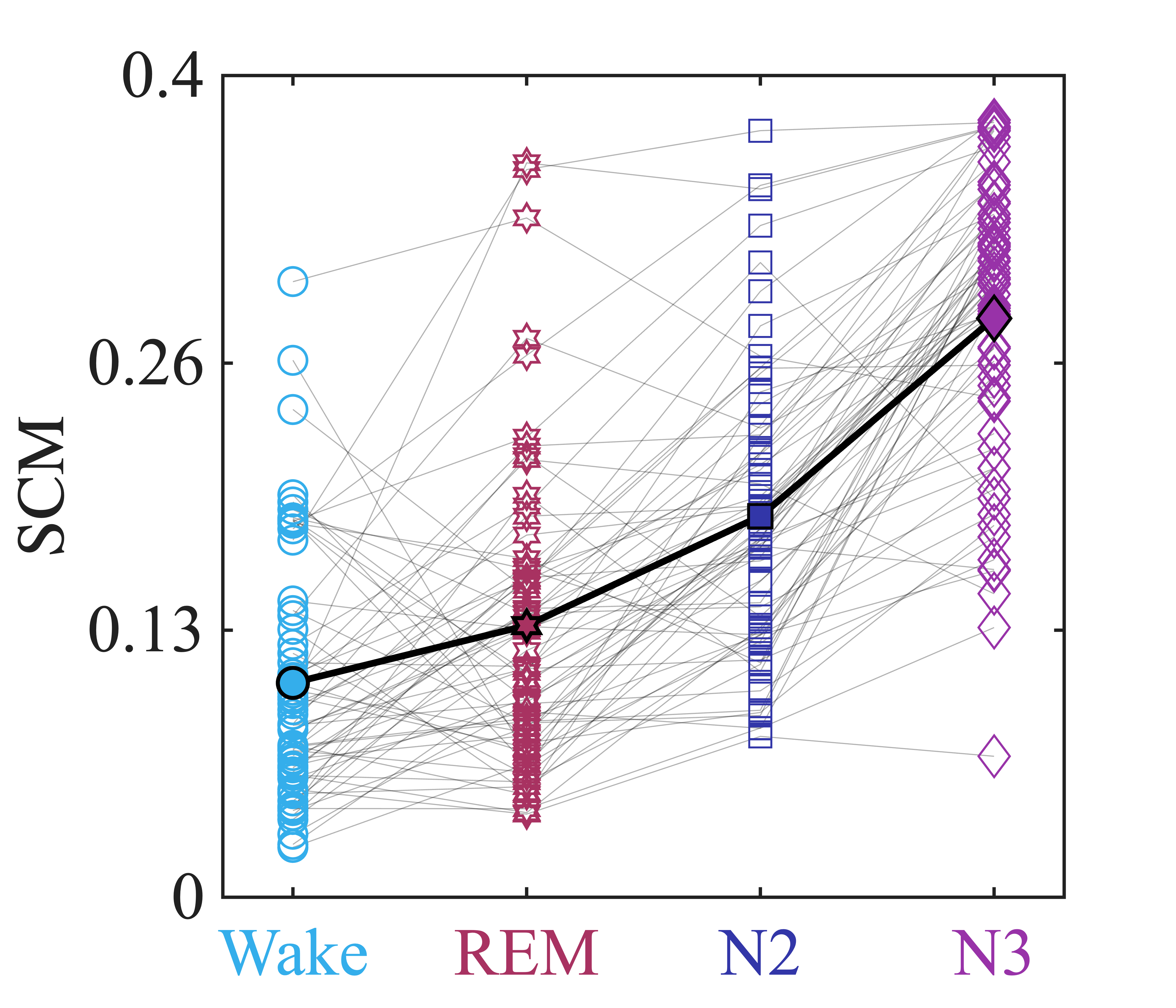}
\end{minipage}
\medskip
\begin{minipage}[t]{0.36\linewidth}
\begin{flushleft}(c)\end{flushleft}
\includegraphics[width=\linewidth]{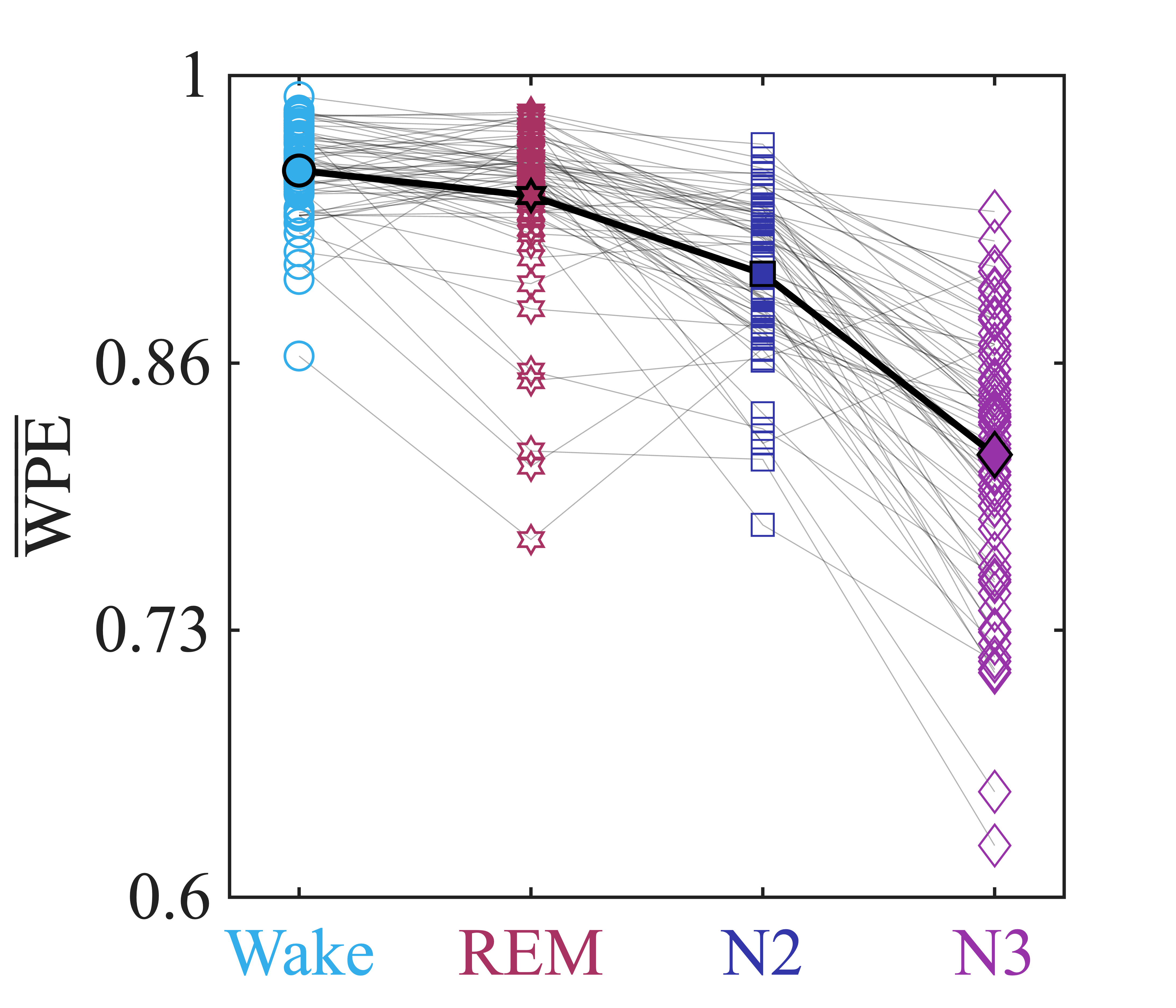}
\end{minipage}
\begin{minipage}[t]{0.36\linewidth}
\begin{flushleft}(d)\end{flushleft}
\includegraphics[width=\linewidth]{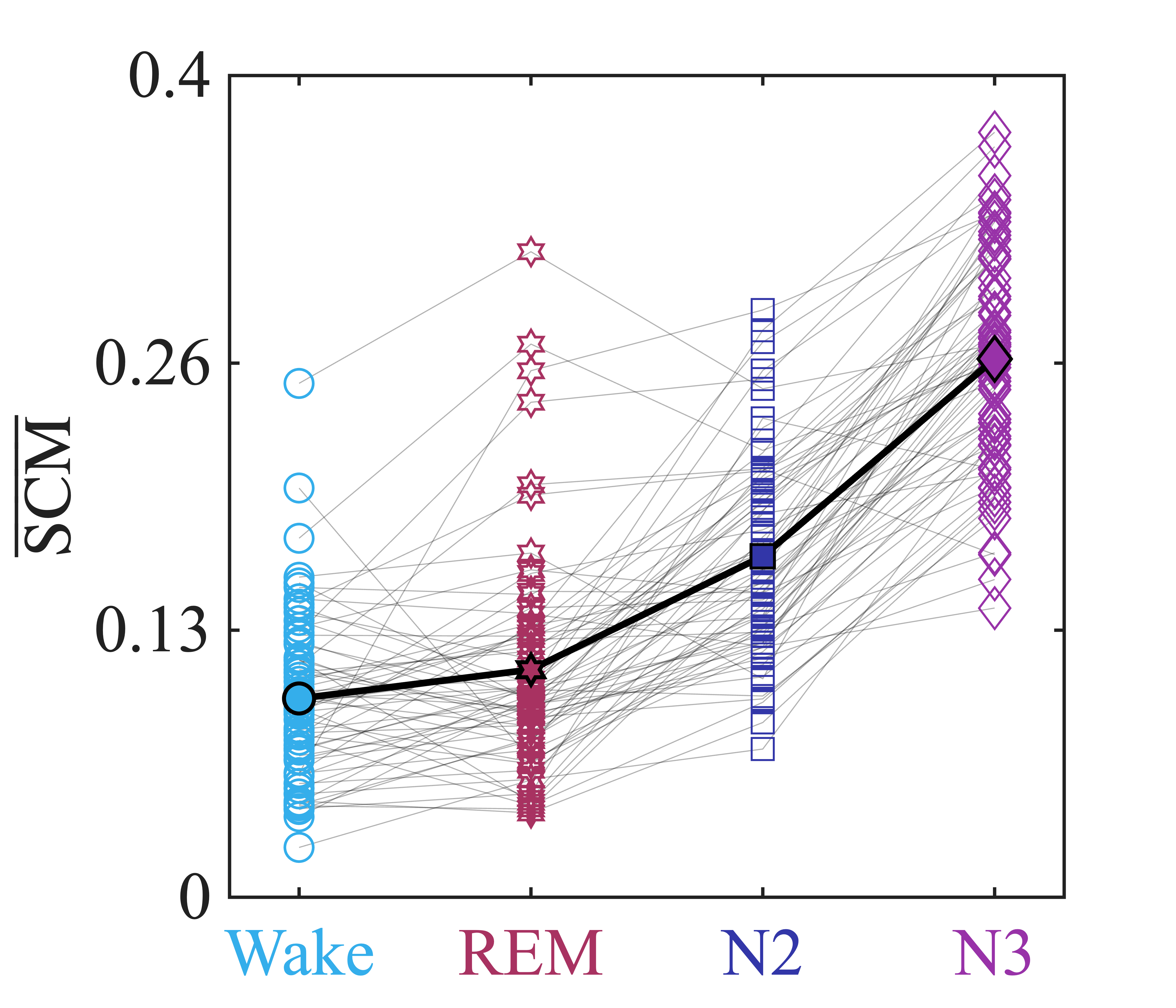}
\end{minipage}
\caption{ Patient-level trajectories of Weighted Permutation Entropy (WPE) and Statistical Complexity Measure (SCM) across the four sleep stages (Wake, REM, N2, and N3) for the 65 participants. (a,b) WPE and SCM, respectively, computed from the time series obtained by averaging the iEEG signals across all valid channels of each patient. (c,d) WPE and SCM, respectively, computed independently for each channel and subsequently averaged across the valid channels of each patient. Gray lines connect the values from the same participant across sleep stages, highlighting the predominant progression from Wake to REM, N2, and N3, characterized by decreasing WPE and increasing SCM.}
\label{fig:caminho_pacientes}
\end{figure}

Initially, for each of the 65 patients, we considered all available time series and computed the average of the iEEG signals across their valid channels. In this way, for each sleep stage, we obtained a single representative time series for each patient, from which WPE and SCM were calculated. This approach allows us to individually track the ``path'' followed by the different sleep stages for each participant, with each trajectory represented by a semi-transparent line in Figure~\ref{fig:caminho_pacientes}. The results are shown in panel (a) for WPE and panel (b) for SCM. Despite the interindividual variability, a well-defined predominant direction can be observed. The Wake $\rightarrow$ REM $\rightarrow$ N2 $\rightarrow$ N3 path, characterized by a progressive decrease in WPE and an increase in SCM, is the most frequent pattern: \novo{31 of the 65 patients ($47.69\%$)} exhibit the ordering Wake $>$ REM $>$ N2 $>$ N3 for WPE, whereas \novo{30 ($46.15\%$)} exhibit the complementary ordering N3 $>$ N2 $>$ REM $>$ Wake for SCM. The second most frequent path corresponds to an inversion only between Wake and REM, namely REM $\rightarrow$ Wake $\rightarrow$ N2 $\rightarrow$ N3. This pattern is observed in \novo{15 patients ($23.07\%$)} for WPE and in \novo{14 ($21.53\%$)} for SCM.

The patients can also be aggregated using the second strategy. In this case, WPE and SCM are first calculated for each individual channel and, subsequently, the resulting measures are averaged across the valid channels of each patient. Thus, for each participant and sleep stage, a pair of values $\overline{\mathrm{WPE}}$ and $\overline{\mathrm{SCM}}$ is obtained. The individual trajectories resulting from this aggregation strategy are shown in Figure~\ref{fig:caminho_pacientes}(c) and (d), for $\overline{\mathrm{WPE}}$ and $\overline{\mathrm{SCM}}$, respectively. Again, the predominant path is Wake $\rightarrow$ REM $\rightarrow$ N2 $\rightarrow$ N3, associated with a decrease in $\overline{\mathrm{WPE}}$ and an increase in $\overline{\mathrm{SCM}}$. In this case, \novo{34 of the 65 patients ($52.30\%$)} follow exactly this ordering for \novo{WPE and 33 ($50.77\%$) for SCM}. The second most frequent pattern is again REM $\rightarrow$ Wake $\rightarrow$ N2 $\rightarrow$ N3, observed in \novo{22 patients ($33.84\%$)} for WPE and in \novo{21 ($32.30\%$)} for SCM. The agreement between the two aggregation strategies shows that the predominant direction of the trajectories does not depend on the order in which channel averaging and the calculation of the quantifiers are performed.

These proportions are particularly remarkable considering that there are 24 possible orderings for four sleep stages. In addition, the most frequent inversions occur mainly between dynamically similar stages, especially Wake and REM, consistent with the proximity of their distributions observed in the previous analyses. The representation of individual trajectories also makes it possible to identify which orderings are rare or virtually absent in the population. In particular, configurations that place N3 at the opposite extreme of the predominant trend, that is, among the states with higher WPE or lower SCM, have zero or very low frequency. This result reinforces that, although individual variations exist in the ordering of the intermediate stages, particularly between Wake and REM, the position of N3 as the state with lower entropy and higher statistical complexity is particularly robust.

As a control analysis, we repeated the characterization of the individual trajectories using conventional Permutation Entropy (PE) and the statistical complexity corresponding to the unweighted ordinal distribution. When the signals from each participant's channels were first averaged and the quantifiers were subsequently computed, the trajectories showed greater heterogeneity than in the WPE analysis. For PE, the most frequent ordering was the same as that observed with WPE (Wake $>$ REM $>$ N2 $>$ N3), but with a \novo{lower percentage: 25 of the 65 patients ($38.46\%$)}. For statistical complexity, the most frequent ordering, also the same as that observed with WPE, namely N3 $>$ N2 $>$ REM $>$ Wake, occurred in also \novo{25 participants ($38.46\%$)}. In the other approach, \novo{in which} the quantifiers were first computed independently for each channel and subsequently averaged within each participant, the organization became more consistent, although it was still less pronounced than with WPE. The most frequent PE trajectory was again Wake $>$ REM $>$ N2 $>$ N3, observed in \novo{31 patients ($47.69\%$)}. For statistical complexity, the trajectory N3 $>$ N2 $>$ REM $>$ Wake occurred in \novo{27 participants ($41.53\%$)}.

This interpretation is consistent with Jeantin et al.~\cite{jeantin2026atlas}, who, using conventional PE on the same dataset, observed high and similar values during Wake and REM, followed by a reduction in N2 and an even more pronounced decrease in N3. Therefore, PE appears to adequately capture the broader reduction in entropy during NREM sleep, particularly in N3. However, our results show that incorporating amplitude weighting in WPE provides a more consistent discrimination among sleep stages, particularly among the intermediate states, and a more stable organization of the complete Wake $\rightarrow$ REM $\rightarrow$ N2 $\rightarrow$ N3 trajectory. In this sense, for characterizing the progression across the four sleep stages, WPE proved to be more robust than conventional PE.

The overlap between the trajectories in Fig.~\ref{fig:caminho_pacientes}, particularly between Wake and REM, does not necessarily imply that the states are indistinguishable, since small shifts may occur in the same direction for most patients. To assess this possibility, we tested the organization of the states directly in the SCM$\times$WPE plane, taking each patient as the unit of observation and representing each state by the point $(\overline{\mathrm{WPE}},\overline{\mathrm{SCM}})$. The two quantities were treated jointly as coordinates of the same point using permutation tests in which state labels were shuffled only within each patient, thereby preserving inter-individual differences. In the global test, the statistic was the sum of the squared Euclidean distances between the centroid of each state and the overall centroid, equivalent to a PERMANOVA with Euclidean distance~\cite{anderson2001new}. For the six pairwise comparisons, the statistic was the Euclidean distance $d$ between centroids, and the null distribution was obtained by randomly reversing each patient's displacement vector $(\Delta\mathrm{WPE},\Delta\mathrm{SCM})$~\cite{pitman1937significance}; $p$-values were corrected using the Holm procedure~\cite{holm1979simple}. All tests used $10^5$ permutations, so that the smallest attainable $p$-value was $\approx10^{-5}$.

In the global test, none of the permutations reproduced the observed separation among the four states ($p\leq1.0\times10^{-5}$), indicating that position in the plane depends systematically on brain state. All six pairwise comparisons remained significant after Holm correction. The closest pair was Wake--REM ($d=\novo{0.019}$; $p_{\mathrm{Holm}}=\novo{0.012}$), for which \novo{39} of 65 patients (\novo{$60\%$}) simultaneously showed a decrease in WPE and an increase in SCM. The remaining comparisons reached the resolution limit of the test ($p_{\mathrm{Holm}}\approx6.0\times10^{-5}$), so the magnitude of the separations is better characterized by the distance between centroids, which increases from REM--N2 ($d=\novo{0.070}$) and Wake--N2 ($d=\novo{0.089}$) to N2--N3 ($d=\novo{0.137}$), REM--N3 ($d=\novo{0.207}$), and Wake--N3 ($d=\novo{0.226}$). Across these five pairs, the fraction of patients showing displacement in the expected direction ranged from $88\%$ to $100\%$, and in the transition from Wake to N3 all 65 patients exhibited lower WPE and higher SCM. The Wake~$\rightarrow$~REM~$\rightarrow$~N2~$\rightarrow$~N3 trajectory therefore represents a shift shared by the vast majority of participants rather than an effect driven by the population average. Its steps, however, are not equally spaced: Wake and REM remain close, whereas the separation increases as NREM sleep deepens.

\novo{These conclusions were not specific to $\tau$ = 8 and $D$ = 6. For delays between 25 and 100 ms and for $D$ = 4, 5, and 6, the global test remained significant, the stage centroids preserved the Wake~$\rightarrow$~REM~$\rightarrow$~N2~$\rightarrow$~N3 ordering, and all pairwise separations involving NREM stages reached the resolution limit of the test (Supplementary Note S2). The Wake–REM separation was the only scale-dependent result. It was significant between approximately 30 and 65 ms, with about 40 of the 65 patients showing lower WPE and higher SCM in REM, but not at longer delays, and the two states exchanged positions at delays shorter than 25 ms (Supplementary Fig. S3)}.

\subsection{Region-level analysis}

\begin{figure}
\centering
\begin{minipage}{0.95\linewidth}
\begin{flushleft}(a)\end{flushleft}
\includegraphics[width=\linewidth]{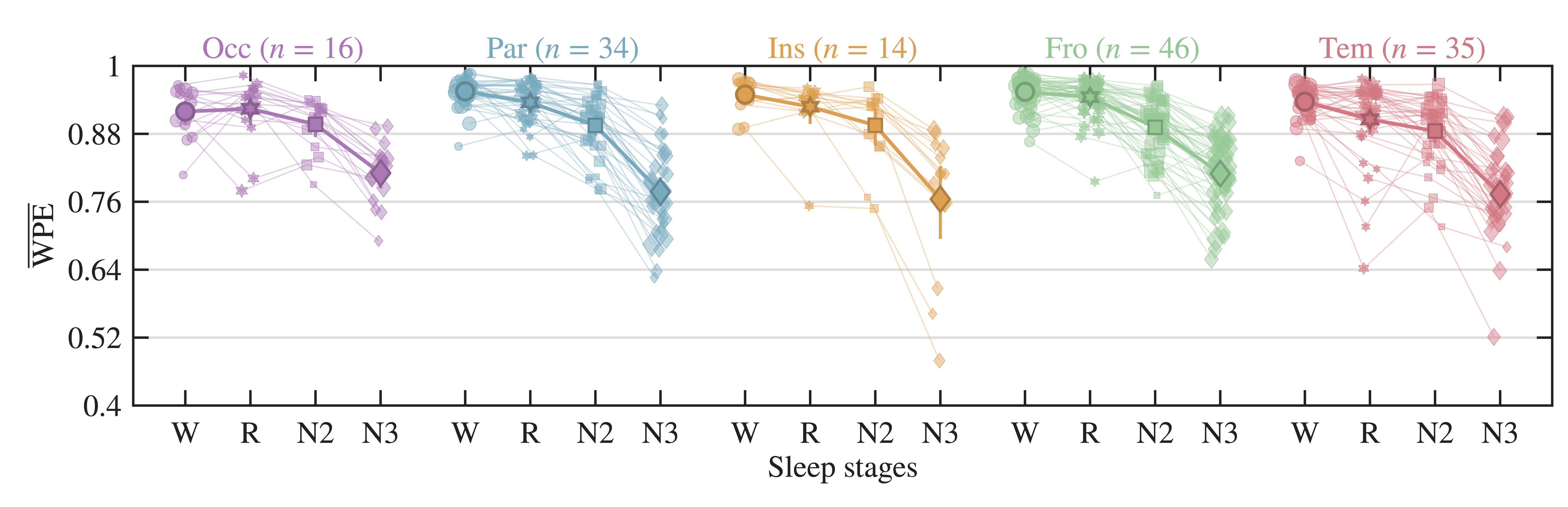}
\end{minipage}
\begin{minipage}{0.95\linewidth}
\begin{flushleft}(b)\end{flushleft}
\includegraphics[width=\linewidth]{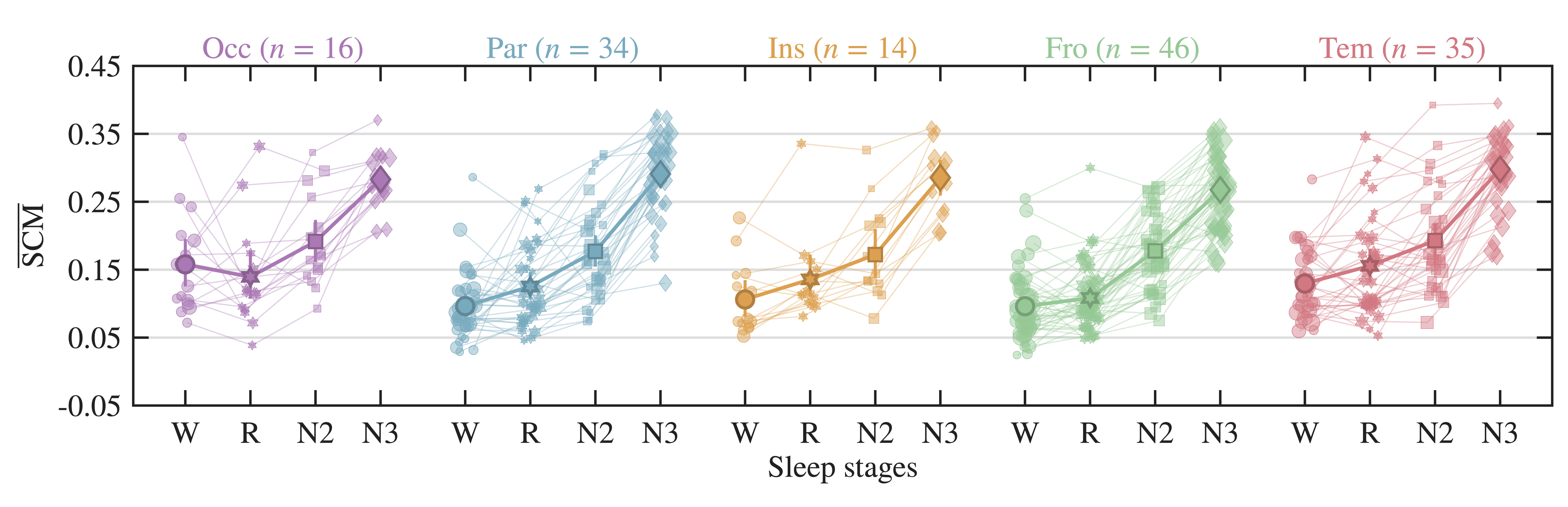}
\end{minipage}
\caption{\novo{Regional \novo{means per patient of} (a) $\overline{\mathrm{WPE}}$ and (b) $\overline{\mathrm{SCM}}$ across Wake (W), REM (R), N2, and N3, computed for each channel with $\tau=\novo{8}$ and $D=6$, \novo{averaged over the channels of each patient within each region,} and grouped into five macro-regions, \novo{namely} occipital (69 channels, 16 patients), parietal (204, 34), insula (49, 14), frontal (469, 46), and temporal (221, 35). \novo{Small markers on the left of each stage tick are single patients, with marker area proportional to the number of channels that contributed, and thin lines join the same patient across stages. Large markers on the right are the means across patients, connected across stages, and vertical bars show 95\% bootstrap confidence intervals obtained by resampling patients ($10^4$ resamples). Marker shapes identify the stages.} Colors follow the regional scheme of Fig.~\ref{fig:position_channels}. In all regions, N3 occupies the lowest-entropy, highest-complexity position, whereas the occipital region departs from the global Wake~$\rightarrow$~REM~$\rightarrow$~N2~$\rightarrow$~N3 ordering in REM \novo{and shows an attenuated displacement in N2}.}}
\label{fig:violino_reg}
\end{figure}

Because sleep is not spatially homogeneous, with slow waves and sleep spindles varying in amplitude, synchronization, and recruitment across different cortical areas~\cite{Nir2011NREM, bernardi2018local, siclari2017local}, we next investigated whether the SCM$\times$WPE plane is traversed in the same way throughout the cortex. Figure~\ref{fig:violino_reg} shows the means per patient and region of (a) $\overline{\mathrm{WPE}}$ and (b) $\overline{\mathrm{SCM}}$ in each of the five macroregions, so that every patient contributes one value per region and stage no matter how many electrodes it has there. We see that, in all regions, N3 occupies the position of lowest entropy and highest complexity, indicating that the endpoint of the global trajectory is reached throughout the cortex. The dispersion across \novo{patients} also follows a common, state-dependent pattern. For $\overline{\mathrm{WPE}}$, the values in Wakefulness and REM are concentrated within a narrow nearby range\novo{, with only a few patients showing lower values in REM,} whereas the distributions broaden in N2 and, especially, in N3, developing long tails toward lower entropy values. For $\overline{\mathrm{SCM}}$, there is already considerable dispersion in Wakefulness and REM\novo{, and it remains large in N3, where the values move toward higher complexity}. This may indicate that deep sleep is simultaneously the most structured and the most heterogeneous state\novo{, at least in terms of entropy,} in agreement with the fact that slow-wave activity manifests with different intensities at different recording locations~\cite{Nir2011NREM, bernardi2018local, siclari2017local}\novo{. The same increase in spread with sleep depth is seen across individual channels in Figs.~\ref{fig:allchannels_subplot_juntos} and~\ref{fig:plano_exp} (b)}.

The regions differ, however, in the path they follow and in the magnitude of their displacement. \novo{This is already visible in Fig.~\ref{fig:violino_reg}, which shows the same means per patient and region used in the statistical tests described below, where each patient contributes equally regardless of electrode coverage.} The parietal, frontal, temporal, and insular regions start from weakly structured wakefulness regimes, with WPE close to one and low SCM. They progress gradually through REM and N2 and reach N3 with comparable displacements. The parietal region, together with the insula and the temporal region, reaches the lowest WPE values and the highest SCM values. The occipital region is the exception. Its wakefulness regime is the most structured among all regions, presenting the lowest WPE and the highest SCM. During REM, it shifts in the direction opposite to the global trend, toward higher WPE values and lower SCM values, and, in N2, it \novo{stays close to} its wakefulness position. A clear displacement appears only in N3, and its magnitude \novo{is about two thirds of that observed in the parietal region (mean displacement in the plane of 0.173 and 0.265, respectively)}. As a result, the occipital \novo{region ends} deep sleep \novo{among the regions with the highest WPE values and the lowest SCM values}, and the occipital and parietal regions reverse their relative positions between Wakefulness and N3. A comparable occipital singularity was reported using sample entropy in the same dataset~\cite{kalamangalam2025gradients}. In that study, occipital areas maintained values similar to those of wakefulness in N2, showed a reduction only in N3, and, in several cases, exhibited higher entropy in REM than in Wakefulness.

We tested these observations using patient--region means as the unit of analysis \novo{(the means shown in Fig.~\ref{fig:violino_reg})}. Region labels were permuted only within each patient, using the 42 patients with recordings in at least two regions ($10^5$ permutations; Holm correction). Within each stage, position in the plane depended on region in Wakefulness ($p_{\mathrm{Holm}}=\novo{2.6\times10^{-3}}$), \novo{was borderline in N3 ($p_{\mathrm{Holm}}=0.055$), and did not depend on region in} REM ($p_{\mathrm{Holm}}=\novo{0.27}$) or N2 ($p_{\mathrm{Holm}}=\novo{0.66}$), indicating that\novo{, at this temporal scale,} regional differentiation is more pronounced at the two extremes of the trajectory\novo{, although it only reaches significance in Wakefulness. The stages at which regional differences are most pronounced depend on the embedding delay (see Supplementary Note~S2)}. We then represented each patient--region combination by its displacement from Wakefulness to REM, N2, and N3 and tested whether these trajectories differed across regions. Indeed, they did ($p=\novo{8.9\times10^{-4}}$). Among all pairwise comparisons, only the occipital--parietal contrast remained significant ($N=14$ patients, $d_{\mathrm{traj}}=\novo{0.122}$, $p_{\mathrm{Holm}}=\novo{0.014}$), and the distance between the two trajectories increased progressively from REM to N2 and N3. \novo{This attenuation was not driven by a small subset of patients with extensive occipital coverage. No patient contributed more than 16\% of the channels of any region, and the occipital displacement from Wakefulness to N3 was shorter than the parietal one in 13 of the 14 patients with recordings in both regions (median ratio of 0.63). The confidence interval of the mean occipital displacement (0.135 to 0.210) does not overlap that of the parietal region (0.232 to 0.297). The mean occipital displacement also varied only between 0.156 and 0.179 when each patient was removed in turn, remaining below that of all other regions (0.220 to 0.271). These patient-level results are shown in Supplementary Note~S3 and Fig.~S5.} Together with the reversal of the occipital--parietal ordering between Wakefulness and N3, this shows that the regions do not differ by a constant displacement in the plane, but by the way they move between states. Therefore, the direction of the global trajectory is preserved throughout the cortex, while its magnitude varies, and the occipital region follows a markedly attenuated path.

\novo{The difference between regional trajectories and the attenuated occipital path were preserved for delays of approximately 45 to 100 ms with the three embedding dimensions, and from 40 ms with $D$ = 5 and from 35 ms with $D$ = 6. At the adopted delay, the occipital and parietal trajectories differed significantly for $D$ = 5 and 6 but not for $D$ = 4. In these ranges, the occipital and parietal contrast remained the only one to survive Holm correction, with a single exception, and the occipital displacement from Wakefulness to N3 was shorter than the parietal one in 11 to 13 of the 14 patients (Supplementary Note~S2, Fig.~S4).}

\color{black}
\subsection{Whole-brain Model}
\label{Sec:Model}

\begin{figure}
    \centering
    \begin{minipage}[t]{0.47\linewidth}
        \begin{flushleft}(a)\end{flushleft}
        \centering
        \includegraphics[width=\linewidth]{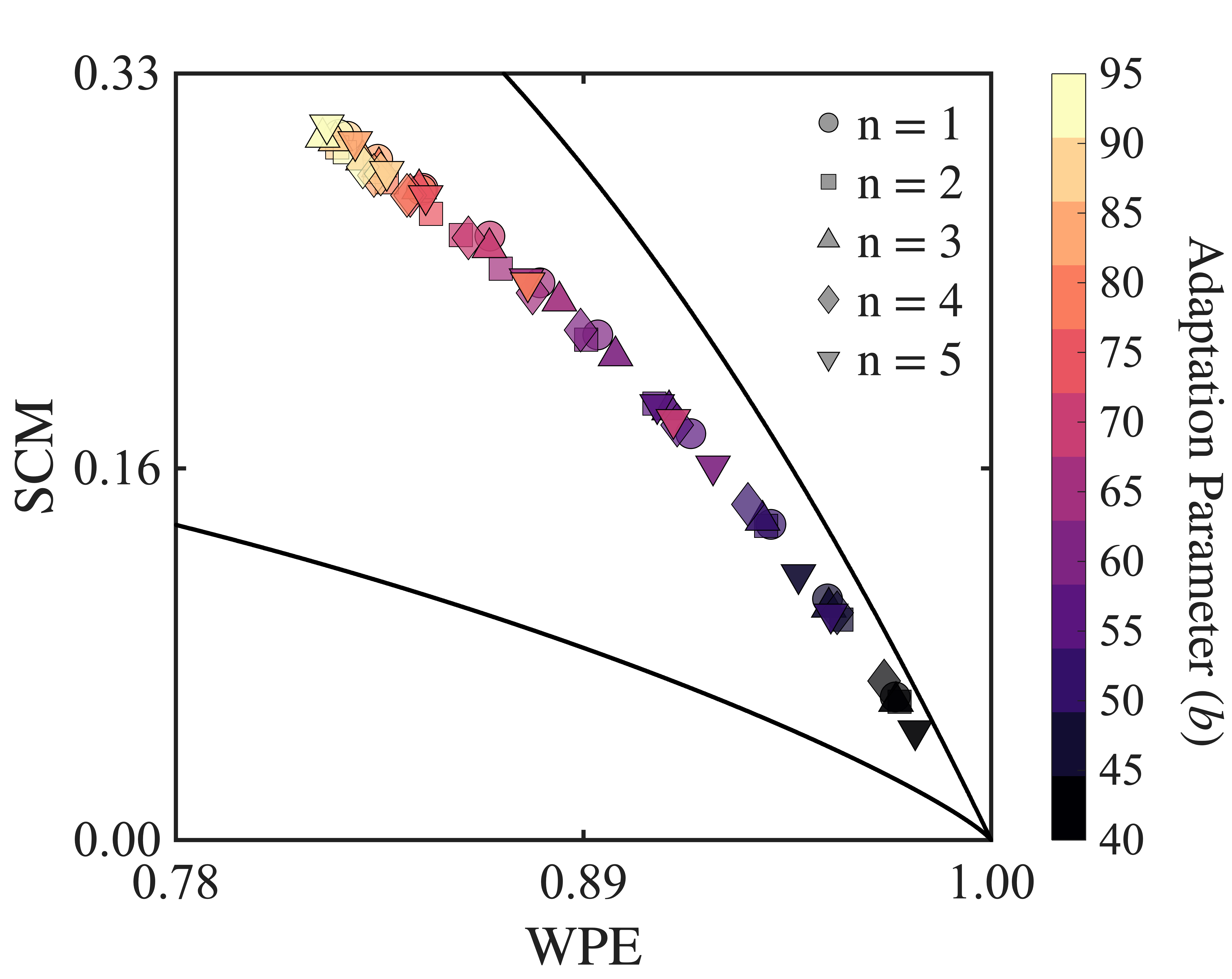}
    \end{minipage}
    \hspace{0.1cm}
    \begin{minipage}[t]{0.47\linewidth}
        \begin{flushleft}(b)\end{flushleft}
        \centering
        \includegraphics[width=\linewidth]{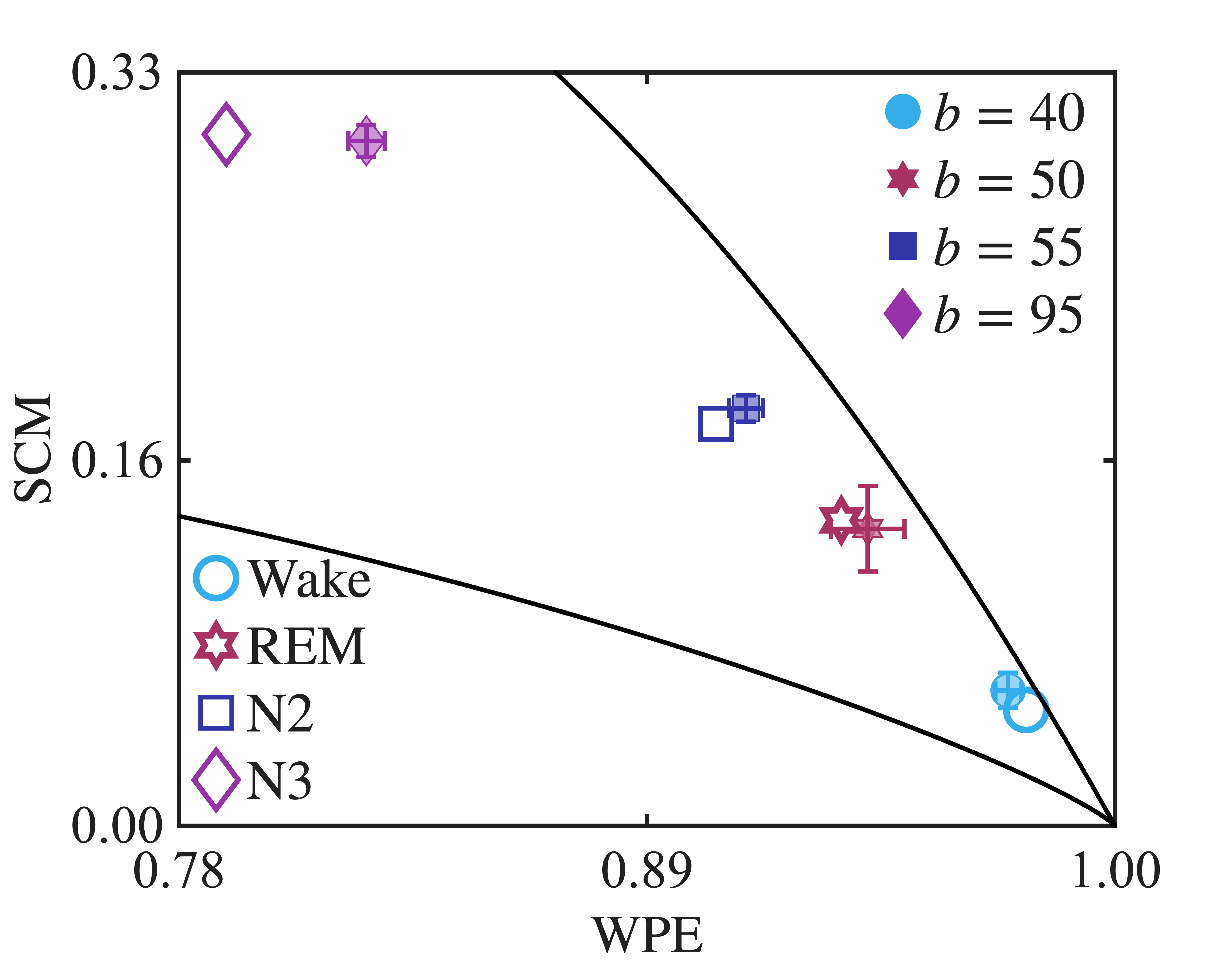}
    \end{minipage}
\caption{\novo{Mapping of simulated brain states in the SCM$\times$WPE plane ($\tau=8$, $D=6$). (a) Systematic variation of the adaptation level $b$ (\novo{40}--95~pA) for five independent noise realizations $n$. Increasing $b$ drove transitions across the plane from high-entropy/low-complexity to low-entropy/high-complexity regimes. (b) Comparison between the empirical sleep states (open symbols) and computational simulations (filled symbols) \novo{at the value of $b$ with the minimum Euclidean distance to each state. Filled symbols and bars are the mean and $\pm1$ standard deviation across the five realizations.} Adaptation levels of \novo{40, 50, 55, and 95}~pA yielded the closest simulated regimes to Wake, REM, N2, and N3, respectively, preserving the overall trajectory observed in the iEEG recordings. Note that the points corresponding to Wake, REM, N2, and N3 have already been shown and discussed in Fig.~\ref{fig:plano_exp}(a).}}
    \label{fig:plane_exp_sim}
\end{figure}

Mechanistically, what could account for the organization of these brain states in the SCM$\times$WPE plane? To address this question, we implemented a human whole-brain computational model capable of reproducing dynamic features of the sleep-wake cycle~\cite{goldman2023simulation, DallaPorta2025WBM}. 
Connectivity among the 68 cortical regions of the Desikan--Killiany parcellation was determined by structural tractography~\cite{schirner2015automated}, and the local dynamics of each region were modeled using the MF-AdEx model~\cite{el2009master,di2019biologically}.
This model incorporates activity-dependent adaptation~\cite{Brette05}, a biophysical mechanism that, when modulated by neuromodulatory input such as ACh, enables transitions between awake-like and sleep-like dynamic regimes~\cite{di2019biologically,dalla2023m,mccormick2020neuromodulation}.

To investigate these transitions and map each state onto the SCM$\times$WPE plane, we systematically varied the adaptation level from $40$ to $95$~pA~\novo{in steps of $5$~pA}, following the same procedure used for the empirical data. 
Specifically, for each adaptation value~\novo{ and for each of five independent noise realizations}, the simulated time series from the 68 cortical regions were averaged, and the corresponding WPE and SCM values were then computed from the resulting mean time series.
This procedure allowed us to assess whether, and for which adaptation values, the model reproduced the organization observed in the empirical data.
As illustrated in Fig.~\ref{fig:plane_exp_sim}(a), increasing adaptation shifted the system across the SCM$\times$WPE plane, from higher entropy and lower complexity at low adaptation values to lower entropy and higher complexity at higher values. 
Notably, the simulated trajectory reproduced the overall direction and ordering observed in the iEEG recordings (Fig.~\ref{fig:plano_exp}).
\novo{This direction was the same in all five realizations (Spearman correlation between $b$ and WPE, $\rho\in[-1.00,-0.92]$; between $b$ and SCM, $\rho\in[0.92,1.00]$). At the values of $b$ closest to each empirical stage (40, 50, 55, and 95~pA), the standard deviation across realizations was at most 0.009 for WPE and 0.019 for SCM, which is smaller than the distance between consecutive empirical stages (0.052--0.171). The variability was larger at intermediate values of $b$ (60--80~pA), where the standard deviation reached 0.024 for WPE and 0.035 for SCM.}

To quantify the correspondence between simulated and empirical states, we computed the Euclidean distance in the SCM$\times$WPE plane between each empirical point and the simulated points~\novo{(mean across realizations)} obtained across the explored adaptation range \novo{(Supplementary Note~S4, Fig.~S6)}. Within this range, the minimum distances were obtained for adaptation levels of $40$, \novo{$50$, $55$}, and $95$~pA for Wake, REM, N2, and N3, respectively, with distances of \novo{$d=0.010$, $0.007$, $0.010$, and $0.033$}. Therefore, in Figure~\ref{fig:plane_exp_sim}(b), these values should be interpreted as the simulated regimes showing the closest correspondence to each empirical state, rather than as unique adaptation values defining the physiological sleep stages.

Together, these results indicate that neuronal adaptation can act as a control parameter driving the modeled system through the SCM$\times$WPE space. This provides a possible mechanistic link between the organization observed in this plane and neuromodulatory changes occurring across the sleep--wake cycle, since neuronal adaptation is regulated by neuromodulatory systems, particularly cholinergic input.

Within the model, the adaptation parameter $b$ governs the post-spike adaptation current in excitatory regular-spiking neurons and has been shown to drive transitions between asynchronous irregular dynamics and slow oscillatory Up and Down dynamics in conductance-based spiking networks~\cite{di2019biologically}. Lower values of $b$ favor desynchronized, irregular activity, whereas increasing $b$ promotes synchronized regimes characterized by slow oscillatory activity~\cite{DallaPorta2025WBM}. This transition is consistent with the known neuromodulatory regulation of cortical dynamics, where wakefulness and REM sleep are associated with high cholinergic tone, which reduces neuronal adaptation, whereas NREM sleep, particularly N3, is associated with reduced cholinergic tone, increased adaptation, and the emergence of slow-wave activity~\cite{lee2012neuromodulation,mccormick2020neuromodulation}.

\section{Conclusion}
\label{Sec:Conclusions}

In this work, we used the SCM$\times$WPE plane to characterize cortical dynamics throughout the sleep--wake cycle using intracranial recordings from humans. The four states followed a clear and consistent trajectory, Wakefulness~$\rightarrow$~REM~$\rightarrow$~N2~$\rightarrow$~N3, with WPE decreasing and SCM increasing as sleep deepened, which suggests that cortical activity becomes less diverse in its ordinal patterns, while the distribution of these patterns becomes progressively more structured.

We also observed that this ordering was not a byproduct of averaging across channels, regardless of how this was done. It persisted when each patient was treated as the unit of analysis, and all four states were significantly separated, including Wakefulness and REM~\novo{ at intermediate temporal scales (approximately 30--65~ms)}, whose separation was small but consistent across patients. \novo{The expected displacement between stages was observed in 88\%--100\% of the patients for all pairs except Wake--REM (60\%).} Deviations from the overall ordering mainly involved an inversion between Wakefulness and REM, whereas N3 was remarkably robust, with all 65 patients showing lower WPE and higher SCM in N3 than in Wakefulness. The ordering was also recovered more consistently by WPE than by conventional permutation entropy (PE), indicating that amplitude weighting improves the characterization of sleep stages. \novo{These results were not specific to the embedding parameters used throughout the main text, the patient-level ordering, the separation of the four states, and the separations involving NREM stages were preserved for delays between 25 and 100~ms and for embedding dimensions $D = 4$, 5, and 6.  Moreover, WPE and SCM are computationally inexpensive, since their cost is linear in the length of the series.}

The subsequent results show that the trajectory is not uniform across the cortex. \novo{The trajectories differed across regions over a broad range of temporal scales, and the occipital--parietal contrast was the only one to survive the Holm correction performed, whereas the stages at which regional differences were most evident depended on the embedding delay.} The parietal, insular, frontal, and temporal regions reproduced the global pattern, whereas the occipital region, in contrast, started from the more structured wakefulness regime, moved in the opposite direction in REM, remained close to its wakefulness position in N2, and shifted clearly only in N3, by \novo{about two thirds of the parietal displacement}. Thus, cortical regions may differ not only in their position in the plane, but also in the path they follow between states, corroborating the literature stating that sleep is a regionally heterogeneous phenomenon.

Finally, the whole-brain model provided a candidate mechanism for this organization. Increasing a single neuromodulation-sensitive parameter, neuronal adaptation $b$, was sufficient to shift the simulated dynamics in the same direction as the empirical states and reproduce their ordering, \novo{in all five independent noise realizations,} with this correspondence being qualitative. The adaptation levels with the best correspondence are not physiological estimates of the individual stages, and a single adaptation axis cannot capture the mechanisms that specifically distinguish REM, such as the withdrawal of aminergic tone. \novo{More generally, because of parameter degeneracy, different combinations of parameters and circuit mechanisms may give rise to similar macroscopic network dynamics~\cite{marder2014neuromodulation,sacha2025computational}. Thus, our results identify neuronal adaptation as a biologically plausible and sufficient mechanism for reproducing the observed organization, but do not imply that it is uniquely responsible for the empirical trajectory.}

These conclusions present some limitations. The recordings come from patients with epilepsy, although they are restricted to channels with probably normal activity, each stage is represented by \novo{54-s} segments, and occipital coverage was limited~\novo{ (16 patients), although the attenuated occipital trajectory was consistent across individual patients}. \novo{In addition, although the main results were robust to the choice of embedding delay and dimension, the Wake--REM separation and the stage-specific regional differences depended on the temporal scale of the ordinal analysis. In the model, only the adaptation level was varied systematically, and the effects of the coupling strength and of the noise amplitude were not explored.} Future work could test whether regionally heterogeneous adaptation or cholinergic modulation reproduces the attenuated occipital trajectory and extend the SCM$\times$WPE framework to anesthesia and sleep disorders. More broadly, our results support neuronal adaptation as a plausible link between cellular mechanisms and the large-scale reorganization of cortical dynamics throughout sleep.

\section*{Data and Code Availability}
The iEEG recordings analyzed in this study are publicly available in the \href{https://mni-open-ieegatlas.research.mcgill.ca/main.php?}{MNI Open iEEG Atlas}. The functions used to compute permutation entropy (PE), weighted permutation entropy (WPE), and the statistical complexity measure (SCM), as well as the code for the whole-brain simulations, are publicly available in the GitHub repository \href{https://github.com/NeuroLab-UFAL/sleep-complexity-entropy-plane}{\texttt{sleep-complexity-entropy-plane}}.

\section*{Supplementary Material}
\novo{The Supplementary Material contains: the number of ordinal vectors and the choice of the embedding delay (Note~S1, Fig.~S1); the robustness to the embedding delay and dimension (Note~S2, Figs.~S2--S4); the patient-level view of the regional analysis (Note~S3, Fig.~S5, Table~S1); and the distance between simulated and empirical states (Note~S4, Fig.~S6).}

\section*{Declaration of generative AI and AI-assisted technologies in the writing process}
During the preparation of this work, the authors used a generative large language model (LLM) to improve the English language of the manuscript. After using this tool, the authors reviewed and edited the content as needed and take full responsibility for the content of the publication.

\section*{Declaration of competing interest}
The authors declare that they have no known competing financial interests or personal relationships that could have appeared
to influence the work reported in this paper.

\section*{Author Contributions}
Helena Bordini de Lucas analyzed the experimental and simulated data and prepared all results for discussion. Leonardo Dalla Porta performed the computational simulations. Fernanda Selingardi Matias conceived and supervised the project. H. B.L., L. D. P., and F. S. M. wrote the manuscript. All authors discussed the results and reviewed the final version of the manuscript.

\section*{Acknowledgments}
The authors thank UFAL, UIB, IDIBAPS, CNPq (Grants No. 402359/2022-4,  314092/2021-8, and 444500/2024-3), FAPEAL (Grant No.
APQ2022021000015), CAPES (Grants 88887.607492/2021-00, and 88881.690510/2022-01), and L’ORÉAL-
UNESCO-ABC For Women In Science (Para Mulheres na
Ciência) for financial support.  This research is also supported
by INCT-NeuroComp (CNPq Grant 408389/2024-9), INFRASLOW PID2023-152918OB-I00 financed by MICIU/AEI/10.13039/501100011033/FEDER, UE, by the European Union (ERC, NEMESIS, project number 101071900) and AGAUR co-funded by Departament de Recerca i Universitats de la Generalitat de Catalunya (AGAUR 2021-SGR-01165). 


\bibliographystyle{unsrtnat}
\bibliography{helena2025}
\includepdf[pages=-]{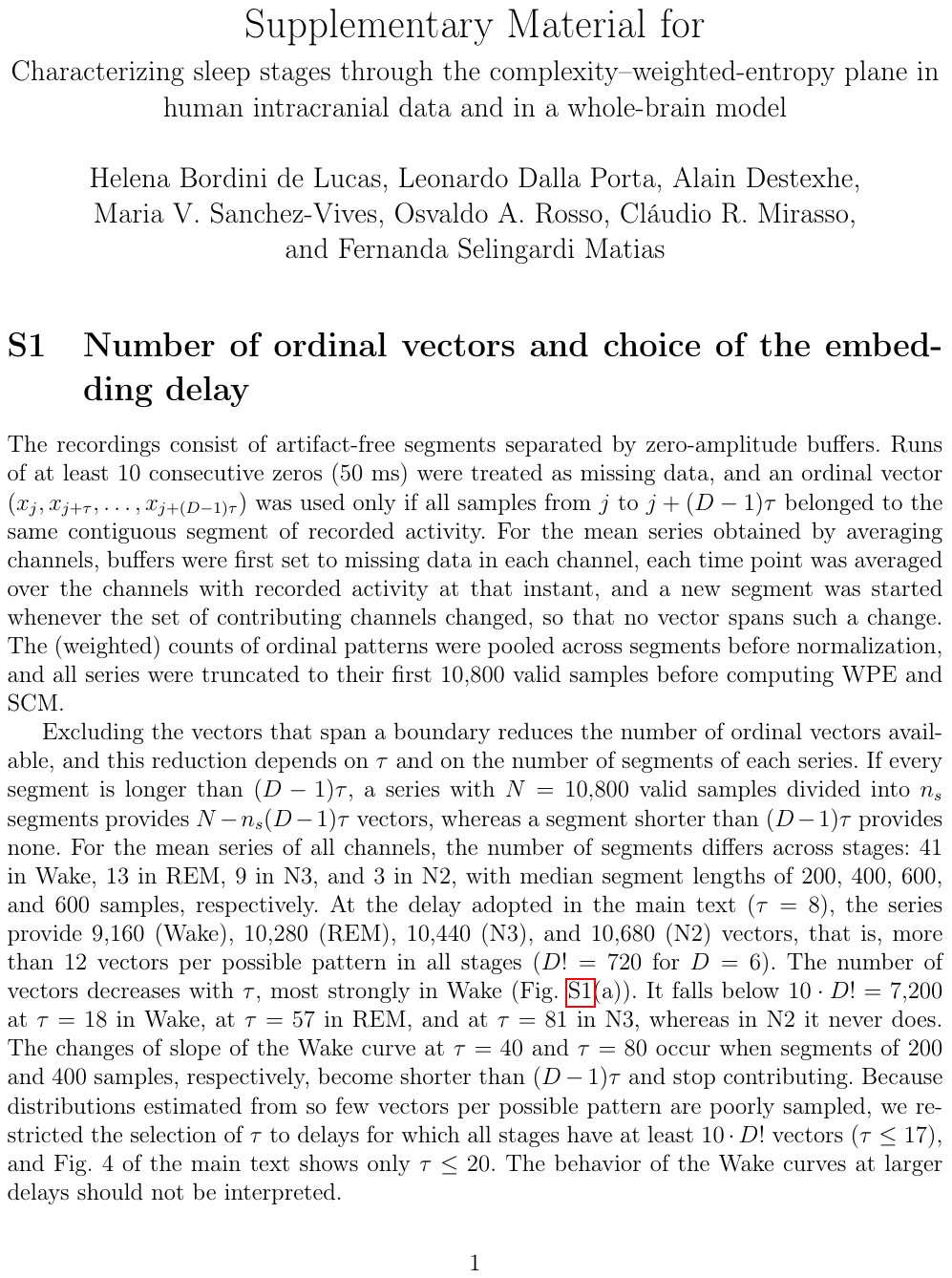}
\end{document}